\documentclass[a4paper,11pt]{article}
\usepackage[T1]{fontenc}      % codifica dei font in uscita
\usepackage[utf8]{inputenc}   % lettere accentate da tastiera
\usepackage[english]{babel}   % lingua principale del documento
\usepackage{url}              % per scrivere gli indirizzi Internet
\usepackage{bm}
\usepackage{amsmath}
\usepackage{amsfonts}
\usepackage{graphicx}
\usepackage[]{appendix}
\usepackage{listings,lstautogobble}

\usepackage{tabularx}
\usepackage{pdflscape}
\usepackage{longtable}
\usepackage{bigints}
\usepackage{amsmath}
\usepackage{amssymb}
\usepackage{float}
\usepackage[small,bf]{caption2}
\usepackage{bm}
\usepackage{geometry}
\usepackage{etoolbox}
\usepackage{amsthm}
\usepackage{amsmath}
\usepackage{xcolor}
\usepackage{algorithm}
\usepackage{algpseudocode}
\usepackage{bm}

\DeclareMathOperator{\EX}{\mathbb{E}}

\usepackage{geometry}
\usepackage{graphicx}
\graphicspath{{img/}}

\usepackage{scalerel}
\usepackage{xcolor}

\usepackage{multirow}

\definecolor{blucite}{RGB}{12,127,172}
\usepackage[authoryear]{natbib}
\usepackage[colorlinks, citecolor=blucite]{hyperref}

\title{\LARGE \textbf{Joint modelling of recurrent and terminal events with discretely-distributed non-parametric frailty:\\ application on re-hospitalizations and death in heart failure patients}\\\vspace{0.4cm}}
\author{\Large  Chiara Masci$^{1,\dagger}$\hspace{1.2cm} Marta Spreafico$^{2,\dagger}$ \hspace{1.4cm} Francesca Ieva$^{1,3}$\\
	\small \texttt{chiara.masci@polimi.it} \quad \texttt{m.spreafico@math.leidenuniv.nl} \quad \texttt{francesca.ieva@polimi.it}\\
 \vspace{-2mm}\\
	\small $^1$MOX Lab, Department of Mathematics, Politecnico di Milano, Milan 20133, Italy\\
	\small $^2$Mathematical Institute, Leiden University, Leiden 2333 CA, The Netherlands\\
	\small $^3$Human Technopole, Health Data Science Center, Milan 20157, Italy\\
 \vspace{1mm}\\ \small
	$^\dagger$\textit{These authors contributed equally to this article.}}

\begin{document}
\normalsize
\maketitle

\begin{abstract}
\small
In the context of clinical and biomedical studies, \textit{joint frailty models} have been developed to study the joint temporal evolution of \textit{recurrent and terminal events}, capturing both the heterogeneous susceptibility to experiencing a new episode and the dependence between the two processes.
While discretely-distributed frailty is usually more exploitable by clinicians and healthcare providers, existing literature on joint frailty models predominantly assumes continuous distributions for the random effects.
In this article, we present a novel \textit{joint frailty model} that assumes \textit{bivariate discretely-distributed non-parametric frailties}, with an unknown finite number of mass points.
This approach facilitates the identification of latent structures among subjects, grouping them into sub-populations defined by a shared frailty value. We propose an estimation routine via Expectation-Maximization algorithm, which not only estimates the number of subgroups but also serves as an unsupervised classification tool.
This work is motivated by a study of patients with Heart Failure (HF) receiving ACE inhibitors treatment in the Lombardia region of Italy. Recurrent events of interest are hospitalizations due to HF and terminal event is death for any cause.\\
\vspace{2mm}
\end{abstract}

\vspace{1mm}
\normalsize
%%%%% INTRODUCTION %%%%%%%%%%%%%%%%%%%%%%%%%%%%%%%%%%%%%%%%%%%%%%%%%%
\section{Introduction}
\label{sec1}

\textit{Recurrent} or \textit{repeated events} are common in many clinical and biomedical studies, as patients usually experience the same event multiple times. Typical situations are follow-up visits, hospital admissions, tumour relapses, heart attacks and many others. 
In the recurrent event framework, classic survival approaches are not suitable as they discard the correlation between subsequent events in the same subject.
A wide literature about recurrent events modelling has hence flourished in past years \citep{surv1,Therneausurvival-book,Cook2007,Aalen2008-wd,amorim2015modelling,ozga2018systematic}. Among others, \textit{frailty models} handle repeated episodes by introducing a random effect which takes a common value for each group of dependent observations \citep{hougaard1995frailty,Hougaard2012-ex,surv1,Therneausurvival-book,Rondeau2003,Rondeau2006,Cook2007,Aalen2008-wd,amorim2015modelling,Gasperoni}.
Given the existence of heterogeneous susceptibility to the risk of recurrent events among subjects, the random term can describe the excess risk or frailty of different individuals, accounting for unexplained heterogeneity not covered by observed covariates. 
Frailty models are particularly well suited to handle hierarchical data structures, such as subjects nested within groups (e.g., patients nested within hospitals).
Nonetheless, an individual's recurrent process duration may be influenced by a terminal event, such as study end, loss to follow-up, or death. Death can prematurely end repeated events, and the terminal event time might be affected by the recurrent event history. Increased occurrences of serious events (e.g., re-hospitalizations) often raise the risk of death, challenging the assumption of independent censoring. As a result, joint frailty models of recurrent and terminal processes has gained significant attention.

\textit{Joint frailty models} analyse both processes over time, treating the terminal process as informative censoring and accounting for their dependence. They capture both the correlation among repeated events and the dependence between repeated and terminal processes by incorporating random effects in both hazard functions.
\cite{jfm_lancaster} initially proposed a parametric model for repeated episodes via Poisson process with a rate function that shares the same subject-specific frailty as the time-to-death hazard, assuming the two processes independent given the frailty term.
\cite{Huang} introduced a joint frailty model for clustered data with informative censoring, sharing a log-normal frailty between censoring and failure rates at the cluster level.
\cite{Liu2004}, \cite{Huang2007}, and \cite{RondeauJFM} considered joint models with shared frailty terms that apply differently for the two hazard functions. \cite{Liu2004} and \cite{Huang2007} used gamma frailty, focusing on time to events (i.e., calendar times) and time between events (i.e., gap times), respectively, and employing a Monte Carlo Expectation–Maximization algorithm for estimation. Alternatively, \cite{RondeauJFM} proposed a non-parametric penalized likelihood estimation method, accommodating gamma and log-normal shared frailty and handling both calendar and gap times. This approach also estimates (smoothed) hazard functions, which often have a meaningful interpretation in epidemiological studies.
\cite{jfm_zeng} generalized joint shared-frailty models using a variety of transformation models, including various possible multivariate random-effects distributions.
Further extensions included Bayesian non-parametric approaches \citep{Paulon}, handling various situations like zero-inflated recurrent events \citep{jfm_liu2016}, nested clustered data for family studies \citep{jfm_Choi2020}, cure fraction models \citep{jfm_cure1,jfm_cure2}, and non-proportional hazards through generalized survival models \citep{chauvet2023}.

Moved by the need for a more flexible approach enabling two correlated random effects to jointly model the dependence between recurrent event and terminal event hazard rates, \cite{NgJFMBiostat} adapted the formulation of \cite{jfm_cure1} to cases without long-term survivors. They focused on patients' gap times between events and developed a joint frailty model using multivariate Gaussian random effects within a generalized linear mixed model. Unlike traditional joint shared-frailty models, this approach uses two sets of random effects to account for both intra-subject correlation in recurrent event times and individual differences in mortality hazard rates. Their formulation efficiently cancels out the unknown baseline hazard functions from the partial likelihood, making the estimation procedures relatively efficient. This method also captures the positive or negative association between recurrent and terminal events, distinguishing the origin of their dependence.

Despite the strengths presented by the aforementioned joint frailty models, the assumption of Gaussian random effects, which is traditionally the most classical approach to model dependency in repeated observations over longitudinal trajectories, poses an edge in the framework of the precision medicine. This is especially the case when the aim is to identify patients profiles  that exhibit different hazards, for whom costumized targeted interventions are needed.
In this perspective, in the last few decades, a more recent branch of the literature is focusing on the treatment of discretely-distributed random effects \citep{aitkin1996general,aitkin1999general,hartzel2001multinomial,azzimonti2013nonlinear,Masci1,masci2021evaluating,masci2022semiparametric}. Most of the work in this direction has been done in the context of mixed-effects regression models for continuous responses (both univariate and multivariate) and other types of responses in the exponential family, but this approach has been recently extended in the survival analysis framework, by the introduction of discretely-distributed frailties
\citep{Caroni2010,Gasperoni,Cancho2020,Cancho2021,Molina2021}.
The main advantage of modelling discrete random-effects is two-fold. First, it provides a new type of treatment and interpretation of the units at the highest level of the hierarchy, that are clustered into latent subpopulations. 
When considering events nested within patients or patients nested within healthcare providers, this approach enables the clustering of patients or healthcare providers, helping to identify characteristics such as long-term and short-term survivors or more and less successful healthcare providers. Second, the non-parametric discrete distribution of random effects allows for a more flexibly modelling of the grouping effect. It avoids imposing any parametric assumption on it and significantly simplifies the computation of the response marginal distribution, thereby avoiding potential integration issues.
Despite the various proposed models and their contributions to the literature, none of them have yet explored the joint modeling of recurrent and terminal events.

Motivated by the aim of developing a tool for profiling patients and identifying latent subpopulations with varying hazard rates, in this article,
we extend the joint frailty model by \cite{NgJFMBiostat} to the discrete-frailty framework.
The  main innovation is assuming a non-parametric bivariate discrete distribution $P^*$ of the random terms with an unknown finite number of mass points. 
In addition to handling both heterogeneous susceptibility to the event risk and informative censoring, this novel approach can detect a latent structure among subjects, grouping them in sub-populations where individuals are characterized by a common frailty value. Frailty values can then be easily translated in a providers’ assessment, resulting more exploitable from an interpretative point of view. 
Our inspiration comes from \cite{Masci1}, where a semi-parametric mixed effect model with a bivariate discretely-distributed non-parametric random term was used to perform an unsupervised classification of school sub-populations based on student performance distributions.
Along with the model, we propose an estimation routine via Expectation-Maximization algorithm \citep{Dempster,EM} and discuss the possible model design choices.
Since the number of subgroups is estimated by the algorithm and is not being known a priori, the proposed model can also be interpreted as an unsupervised classification tool, encouraging exploration of similarities within grouped subjects and differences between sub-populations.

The approach developed in this paper is motivated by a study on patients with Heart Failure (HF) hospitalized in the Lombardia region of Italy.
HF disease often leads to recurrent hospitalization events \citep{Kennedy2001,Baraldo2013,Rogers2016,Paulon,Spreafico2021BJ2}, which usually herald a substantial worsening of patient's survival prognosis and are terminated by death. In this context our novel approach is hence of interest for several reasons.
First, it addresses informative censoring and detects sub-populations of HF patients at different risks, representing a more informative interpretation tool for medical practice.
It also enables the investigation of the impact of patient-specific (time-varying) characteristics on hospital readmissions and/or mortality.
Specifically, a key component of patient’s care is adherence to medication, that is the process by which patients take their medication as prescribed \citep{Vrijens2012}. 
Proper medication adherence in HF patients can improve clinical outcomes and prevent hospitalization and reduce mortality \citep{Ponikowski2016}.
Specifically, we focus on HF patients receiving Angiotensin-Converting Enzyme (ACE) inhibitors, i.e., disease-modifying drugs of routine use for HF therapy \citep{McMurray2012,Yancy2013,Ponikowski2016}, and we study the effect of time-varying adherence \citep{Spreafico2021BJ1} to ACE therapy on both re-hospitalisations and death. Both the model and its application to HF are innovative contributions to the literature.

The remainder of the article is organized as follows. 
In Section \ref{sec:met}, we introduce notation, joint frailty models and the main novelty of this work, i.e., the \textit{Joint Model with Discretely-distributed non-parametric Frailty (JMDF)} for modelling recurrent and terminal events. Section \ref{sec:est} covers the estimation procedure through Expectation-Maximization algorithm and model design options. In Section \ref{sec:app}, we apply the proposed methodology to the HF administrative database provided by \textit{Regione Lombardia - Healthcare Division} \citep{RegioneLombardia2012}, comparing the results to the joint frailty model by \cite{RondeauJFM} and by \cite{NgJFMBiostat}.
The concludin  Section \ref{sec:discussion} discusses the approach's strengths, limitations, and potential future directions.
Statistical analyses were performed in the \texttt{R} software environment \citep{Rsoftware}. In order to enhance reproducibility and validation of the research, source code is available  at {\url{https://github.com/mspreafico/JMDF}}. 

%%%%% METHODS %%%%%%%%%%%%%%%%%%%%%%%%%%%%%%%%%%%%%%%%%%%%%%%%%%
\section{Methods}
\label{sec:met}
 
In Section \ref{sec:met_notation} we introduce the notation for recurrent and terminal time-to-events considering \textit{gap times}, i.e., times between consecutive events, and we briefly recall the joint frailty modeling proposed by \cite{RondeauJFM} and \cite{NgJFMBiostat}. In section \ref{sec:met_our}, we introduce our novel JMDF methodology. 

\subsection{Notation and state-of-the-art methods} \label{sec:met_notation}

\subsubsection{Notation}\label{sec:met_notationvera}
Let us consider a cohort of $N$ independent individuals, denoted by index $i$ ($i=1,...,N$), in which each subject experiences $J_{i}$ recurrent events, denoted by index $j$. For each subject $i$, let $T^{R}_{ij}$ denote the gap time of recurrent event $j$ with realization $t_{ij}^R$, and $T^{D}_{i}$ the gap time from the last recurrent event to the terminal one (e.g., death) with realization $t_{i}^D$. Both event-times are subject to right censoring, whose time is denoted by $C_{i}$. For each subject $i$ we hence observe $n_i$ gap times, with $n_i=J_i+1$.
Define $T_{ij}=\min\left\{T^{R}_{ij},T^{D}_{i},C_{i}\right\} \, \forall j=1,...,n_i$ as the sequence of random variables defining observed times with realizations $t_{ij}$. Let $\delta^{R}_{ij}$ be the corresponding censoring variables for the recurrent event (1 if $T_{ij}=T^{R}_{ij}$, 0 otherwise), and $\delta^{D}_{i}$ be the censoring variable for the terminal event (1 if $T_{in_{i}}=T^{D}_{i}$, 0 otherwise). Let $\bm{X}^{R}_{ij}\in\mathbb{R}^{p_1}$ denotes the $p_1$-dimensional vector of (fixed or time-dependent) covariates associated to recurrent event $j$ for subject $i$, and $\bm{X}^{D}_{i}\in\mathbb{R}^{p_2}$ be the $p_2$-dimensional vector of covariates associated to its terminal event. The observable data for each patient $i$ are given by $\bm{\mathcal{O}}_i=\left\{ \left(T_{ij},T_{ij}^R,\delta^R_{ij},\boldsymbol{X}_{ij}^R,T_{i}^D,\delta^D_i,\boldsymbol{X}_{i}^D\right); j=1,\dots,n_i\right\}$. The overall set of observable gap times, censoring variables and covariates is denoted by $\bm{\Theta} = \left\{\bm{\mathcal{O}}_i; i=1,\dots,N \right\}.$

\subsubsection{Joint frailty model with shared log-normal  random effects \citep{RondeauJFM}}\label{sec:met_sa_rondeau}
To account for heterogeneity in the data due to unobserved covariates, \cite{RondeauJFM} proposed a joint model that included a common frailty term to the individuals for the two rates related to recurrent and terminal events. Such term is assumed to follow either a Gamma or a log-Normal distribution and acts differently on the two hazard rates by means of a parameter $\alpha$. In particular, the hazard functions for the \textit{joint log-Normal frailty model} are defined by:
\begin{equation}
    \label{rondeauEqNorm}
    \begin{cases}
    h^{R}_{ij}\left(t|\eta_{i},\bm{X}^{R}_{ij}\right) =h^{R}_{0}(t)\exp\left(\bm{\beta}^{T}\bm{X}^{R}_{ij}+\eta_{i} \right) \ \ \\
    h^{D}_{i}\left(t|\eta_{i},\bm{X}^{D}_{i}\right) =h^{D}_{0}(t)\exp\left(\bm{\gamma}^{T}\bm{X}^{D}_{i}+\alpha\eta_{i}\right)
    \end{cases}
\end{equation}
where $h^{R}_{0}(\cdot)$ and $h^{D}_{0}(\cdot)$ are the baseline hazard functions for recurrent and terminal events, respectively; $\bm{\beta}$ and $\bm{\gamma}$ are the vectors of fixed-effect coefficients associated to recurrent and terminal events, respectively. The random effects are iid and distributed as $\eta_{i}\overset{iid}{\sim}\mathcal{N}(0,\sigma^2)$. The parameter $\alpha$ determines direction of the association (if significant) between the two processes.

Parameter estimation is based on a semiparametric penalized likelihood estimation or parametric estimation on the hazard function (see \cite{RondeauJFM} for further details) and it is implemented in the R package \texttt{frailtypack} \citep{frailtypack,frailtypack2017}. 

\subsubsection{Joint frailty model with multivariate Gaussian random effects \citep{NgJFMBiostat}}\label{sec:met_sa_ng}

Recently, \cite{NgJFMBiostat} proposed a joint frailty model for recurrent and terminal events with multivariate Gaussian random effects. The authors denote by $\boldsymbol{u}=\left(u_1,\dots,u_N\right)^T$ and $\boldsymbol{v}=\left(v_1,\dots,v_N\right)^T$
the $N$-dimensional random vectors of $u_i$ and $v_i$ that represent the frailty for the $i$-th subject to account for intra-subject correlation of gap times of the recurrent events and individual differences in mortality hazard rate for the gap time from the last recurrent event to death, respectively.
Their joint frailty model is defined as: 
\begin{equation}
    \label{ngEq}
    \begin{cases}
h^{R}_{ij}\left(t^{R}_{ij}|u_{i},\bm{X}^{R}_{ij}\right) =h^{R}_{0}\left(t^{R}_{ij}\right)\exp\left(\bm{\beta}^{T}\bm{X}^{R}_{ij} + u_{i}\right)\\
    h^{D}_{i}\left(t^{D}_{i}|v_{i},\bm{X}^{D}_{i}\right) =h^{D}_{0}\left(t^{D}_{i}\right)\exp\left(\bm{\gamma}^{T}\bm{X}^{D}_{i} + v_{i}\right)
    \end{cases}
\end{equation}
where $h^{R}_{0}(\cdot)$ and $h^{D}_{0}(\cdot)$ are the baseline hazard functions for recurrent events and death, respectively; $\bm{\beta}$ and $\bm{\gamma}$ are the vectors of fixed-effect coefficients associated to recurrent and terminal events, respectively. The frailty joint distribution is
\begin{equation} \label{eq:Ngform}
    [u_{i},v_{i}] \sim \mathcal{N}_{2}(\bm{0},\bm{\Sigma}) \qquad \qquad \text{with} \qquad \qquad \bm{\Sigma}=\begin{bmatrix}
\theta_{u}^{2} & \rho\theta_{u}\theta_{v}\\
\rho\theta_{u}\theta_{v} & \theta_{v}^{2}  
\end{bmatrix}.
\end{equation}
This formulation allows a well-defined, straightforward interpretation of all the parameters involved, being $\theta^2_u$ and $\theta^2_v$ the quantifiers of unobserved heterogeneity in the two processes, whereas $\rho$ models the dependence between $\bm{u}$ and $\bm{v}$.

Parameter estimation is performed by maximize the sum of two components: (i)  the usual Cox partial log-likelihood of failure times assuming $\bm{u}$ and $\bm{v}$ fixed, i.e., $\ell_1\left(\bm{\beta},\bm{\gamma}|\bm{u},\bm{v}\right)$; (ii) the logarithm of the joint probability density function of random effects $\bm{u}$ and $\bm{v}$, i.e., $\ell_2\left(\bm{u},\bm{v}|\theta_{u},\theta_{v},\rho\right)$. See \cite{NgJFMBiostat} for further details.

\subsection{Joint Model with Discretely-distributed non-parametric Frailty (JMDF)} \label{sec:met_our}

In devising our original methodology, we build upon the framework proposed in \cite{NgJFMBiostat}, while introducing a bivariate non-parametric discrete distribution for the random effects. This choice is driven by the fact that a discrete distribution of frailties not only offers an additional layer of interpretation in medical practice but also provides the opportunity to discover and analyze latent partitions within the cohort of patients under consideration.

The hazard functions in the JMDF for recurrent and terminal events are defined as: 
\begin{equation}
    \label{eq:ourJM}
    \begin{cases}
h^{R}_{ij}\left(t^{R}_{ij}|u_{i},\bm{X}^{R}_{ij}\right) =h^{R}_{0}\left(t^{R}_{ij}\right)\exp\left(\bm{\beta}^{T}\bm{X}^{R}_{ij} + u_{i}\right)\\
    h^{D}_{i}\left(t^{D}_{i}|v_{i},\bm{X}^{D}_{i}\right) =h^{D}_{0}\left(t^{D}_{i}\right)\exp\left(\bm{\gamma}^{T}\bm{X}^{D}_{i} + v_{i}\right)
    \end{cases}
\end{equation}
where $h^{R}_{0}(\cdot)$ and $h^{D}_{0}(\cdot)$ are the baseline hazard functions for recurrent and terminal events, respectively; $\bm{\beta}$ and $\bm{\gamma}$ are the vectors of fixed-effect coefficients associated to recurrent and terminal events, respectively. 
Random effects $u_{i}$ and $v_{i}$ are assumed to be distributed according to an unknown discrete measure with a finite support in $\mathbb{R}^2$, called $P^{*}$: 
\begin{equation}\label{eq:Pstar}
    [u,v]_{i} \overset{iid}{\sim} P^{*} \ \ \ \forall i=1,...,N.
\end{equation}
$P^{*}$ can be characterized by a vector $\bm{\mathcal{P}}=\left(\bm{P}_1,\bm{P}_2,\dots,\bm{P}_K\right)$ of $k=1,\dots,K$ points $\bm{P}_k=\left[P_k^{u},P_k^{v}\right]\in\mathbb{R}^2$  where $K$ is unknown a priori, and a vector $\bm{w}=\left(w_1,w_2,\dots,w_K\right)$ of relative weights. Notice that $\sum_{k=1}^K w_k = 1$ and each weight $w_{k}=\Pr\left([u,v]_{i}=\bm{P}_{k}\right)$ expresses the probability for each subject $i$ (that is the second-level unit) to be assigned to a certain point $k$.

\subsubsection{Likelihood construction.}\label{sec:met_our_lik}
In order to define the likelihood, we initially consider $K$ as fixed and we introduce a set of auxiliary random variables. For each subject $i$, we define an auxiliary random vector $\bm{z}_{i}$ as follows
\begin{equation}
    \bm{z}_{i}=[z_{i1} \ z_{i2} \ z_{i3} \ ... \ z_{iK}] 
\qquad \text{where} \qquad
    z_{ik} = 
   \begin{cases}
    1 \quad &\text{if } [u,v]_{i}=\bm{P}_{k} \\
    0 \quad &\text{otherwise.}
    \end{cases}
\end{equation}
Thus, each auxiliary vector is distributed according to a multivariate Bernoulli distribution of parameters $\bm{w}$.

Assuming that we have observed the realizations of such auxiliary random vectors (collected in the random matrix $\bm{\mathcal{Z}}$), we can express the full likelihood of the model as
\begin{equation}
\mathcal{L}\left(\bm{\Omega};\bm{\Theta}|\bm{\mathcal{Z}}\right)=\prod_{k=1}^{K}\prod_{i=1}^{N} \big[w_k \cdot \mathcal{L}_{ik}(\bm{\Omega};\bm{\mathcal{O}}_{i}|\bm{z}_i)\big]^{z_{ik}}
\end{equation}
where $\bm{\Theta}$ is the overall set of observable data (see Section \ref{sec:met_notation}), and $\bm{\Omega}=\left[\bm{\beta},\bm{\gamma},H_{0}^{R}(t),H_{0}^{D}(t),\bm{w},\bm{\mathcal{P}}\right]$ denotes the quantities to be estimated, with $H_{0}^{R}(t)$ and $H_{0}^{D}(t)$ being the cumulative baseline hazards related to the two processes.
Each individual contribution to the likelihood can be written as the following product
\begin{equation}  \mathcal{L}_{ik}\left(\bm{\Omega};\bm{\mathcal{O}}_{i}|\bm{z}_{i}\right)=\prod_{j=1}^{n_{i}} \mathcal{L}_{ijk}^{R}\left(\bm{\Omega};\bm{\mathcal{O}}^{R}_{ij}\big|\bm{z}_{i}\right)\cdot  \mathcal{L}_{ik}^{D}\left(\bm{\Omega};\bm{\mathcal{O}}^{D}_{i}\big|\bm{z}_{i}\right)
\end{equation}
with
\begin{align}   
 \mathcal{L}_{ijk}^{R}\left(\bm{\Omega};\bm{\mathcal{O}}^{R}_{ij}\big|\bm{z}_{i}\right)=\left[ h_{0}^{R}\left(t_{ij}^{R}\right) \exp \left( \bm{\beta}^{T}\bm{X}^{R}_{ij}+P_{k}^{u}\right)\right]^{\delta_{ij}^{R}} \cdot \exp \left\{-H_{0}^{R}\left(t_{ij}^{R}\right)\exp\left(\bm{\beta}^{T}\bm{X}^{R}_{ij}+P_{k}^{u}\right)\right\}
\end{align}
and
\begin{align}
\mathcal{L}_{ik}^{D}\left(\bm{\Omega};\bm{\mathcal{O}}^{D}_{i}\big|\bm{z}_{i}\right)=\left[ h_{0}^{D}\left(t_{i}^{D}\right) \exp\left(\bm{\gamma}^{T}\bm{X}^{D}_{i}+P_{l}^{v}\right)\right]^{\delta_{i}^{D}} \cdot \exp{\left\{-H_{0}^{D}\left(t_{i}^{D}\right)\exp\left(\bm{\gamma}^{T}\bm{X}^{D}_{i}+P_{k}^{v}\right)\right\}}, 
\end{align}
where $P_{k}^{u}$ and $P_{k}^{v}$ stand for the coordinates of the chosen support point $\bm{P}_{k}$, respectively. 

We can then express the log-likelihood $\ell\left(\bm{\Omega};\bm{\Theta}|\bm{\mathcal{Z}}\right)$ as the sum of three terms:
\begin{equation}
\label{eq:logl}
\ell\left(\bm{\Omega};\bm{\Theta}|\bm{\mathcal{Z}}\right)=
\ell_{w}\left(\bm{\Omega}_{w};\bm{\Theta}|\bm{\mathcal{Z}}\right)+
\ell_{R}\left(\bm{\Omega}_{R};\bm{\Theta}_R|\bm{\mathcal{Z}}\right)+
\ell_{D}\left(\bm{\Omega}_{D};\bm{\Theta}_D|\bm{\mathcal{Z}}\right)
\end{equation}
where 
\begin{equation}
\label{eq:logl_lw}
\ell_{w}\left(\bm{\Omega}_{w};\bm{\Theta}|\bm{\mathcal{Z}}\right)=\sum_{k=1}^{K}\sum_{i=1}^{N}z_{ik}\cdot\log(w_{k}),
\end{equation}
\begin{align}
\label{eq:logl_lr}  \ell_{R}\left(\bm{\Omega}_{R};\bm{\Theta}_R|\bm{\mathcal{Z}}\right)=\sum_{k=1}^{K}\sum_{i=1}^{N}z_{ik}\cdot\Bigg[ & \sum_{j=1}^{n_{i}} \delta_{ij}^{R}\left\{\log\left(h_{0}^{R}\left(t_{ij}^{R}\right)\right)+\bm{\beta}^{T}\bm{X}_{ij}^{R}+P_{k}^{u}\right\} \\ & -H_{0}^{R}\left(t_{ij}^{R}\right)\exp\left(\bm{\beta}^{T}\bm{X}_{ij}^{R}+P_{k}^{u}\right)\Bigg], \nonumber
\end{align}
\begin{align}
\label{eq:logl_ld}
\ell_{D}\left(\bm{\Omega}_{D};\bm{\Theta}_D|\bm{\mathcal{Z}}\right)=\sum_{k=1}^{K}\sum_{i=1}^{N}z_{ik}\cdot\Bigg[ &  \delta_{i}^{D}\left\{ \log\left(h_{0}^{D}\left(t_{i}^{D}\right)\right)+\bm{\gamma}^{T}\bm{X}_{i}^{D}+P_{k}^{v}\right\} \\ & -H_{0}^{D}(t_{i}^{D})\exp\left(\bm{\gamma}^{T}\bm{X}_{i}^{D}+P_{k}^{v}\right)\Bigg]. \nonumber
\end{align}
Estimators for $\bm{\Omega}=[\bm{\Omega}_{w},\bm{\Omega}_{R},\bm{\Omega}_{D}]$ can be obtained by maximizing Eq. \eqref{eq:logl} using the Expectation-Maximization algorithm \citep{Dempster,EM} proposed in Section \ref{sec:est_EM}.

%%%%% ESTIMATION & MODEL DESIGN %%%%%%%%%%%%%%%%%%%%%%%%%%%%%%%%%%%%%%%%%%%%%%%%%%
\section{Estimation and Model design}
\label{sec:est}
In this section we present the Expectation-Maximization algorithm (Section \ref{sec:est_EM}) we propose to estimate parameters, discussing details on the identification of the support points (Section \ref{sec:est_supportReduction}).

\subsection{A tailored Expectation-Maximization algorithm}\label{sec:est_EM}
The log-likelihood in Eq. \eqref{eq:logl} is defined conditionally on the auxiliary random matrix $\bm{\mathcal{Z}}$. In order to maximize it,  we propose a novel Expectation-Maximization algorithm \citep{Dempster,EM, Gasperoni} to estimate $\bm{\Omega}=[\bm{\Omega}_{w},\bm{\Omega}_{R},\bm{\Omega}_{D}]$ for a given number of support mass points $K$.\\

\textbf{Parameter initialization:} The initial step involves parameter initialization, i.e., determining $\bm{\Omega}^{(0)}$. Firstly, the grid of support points for the discrete distributions is initialized to obtain $\left[\bm{\mathcal{P}},\bm{w}\right]^{(0)}$ following the procedure outlined in Section \ref{sec:est_supportReduction}.
Next, two Cox-type models are fitted: one for the recurrent events with $\bm{\mathcal{P}}_u^{(0)}$ as the offset, and another for the terminal event with $\bm{\mathcal{P}}_v^{(0)}$ as the offset. The estimated parameters from these models, along with their corresponding estimated cumulative baseline hazard functions, are then used to initialize the remaining parameters, namely  $\left[\bm{\beta},\bm{\gamma},H_0^R(t),H_0^D(t)\right]^{(0)}$.\\

\textbf{E-step:} At each iteration, the Expectation step consists of computing: 
\begin{equation}    
Q\left(\bm{\Omega}\right) =
\EX_{\bm{\mathcal{Z}}|\hat{\bm{\Omega}}}\left[\ell(\bm{\Omega};\bm{\Theta})\right] =\EX_{\bm{\mathcal{Z}}|\hat{\bm{\Omega}}}\left[ \ell_{w}\left(\bm{\Omega}_{w};\bm{\Theta}\right)\right] +
\EX_{\bm{\mathcal{Z}}|\hat{\bm{\Omega}}}\left[ \ell_{R}\left(\bm{\Omega}_{R};\bm{\Theta}_R\right)\right] +
\EX_{\bm{\mathcal{Z}}|\hat{\bm{\Omega}}}\left[ \ell_{D}\left(\bm{\Omega}_{D};\bm{\Theta}_D\right)\right] 
\end{equation}
which is the expectation over $\bm{\mathcal{Z}}$, given the current estimates of parameters $\hat{\bm{\Omega}}$, of the log-likelihood in Eq. \eqref{eq:logl} for the observed data $\bm{\Theta}$. This reduces to the computations of $\EX\left[z_{ik}\big|\bm{\hat{\Omega}},\bm{\mathcal{O}}_i\right]$ which we indicate as $\mathbb{Z}_{ik}$ and can be derived in closed form using Bayes' theorem:
\begin{equation}
    \mathbb{Z}_{ik}= \small \frac{w_{k}\exp\big\{\sum_{j=1}^{n_{i}}\big(\delta_{ij}^{R}P_{k}^{u}-H_{0}^{R}(t_{ij}^{R})\exp\{\bm{\beta}^{T}\bm{X}_{ij}^{R}+P_{k}^{u}\}\big) + \delta_{i}^{D}P_{k}^{v}-H_{0}^{D}(t_{i}^{D})\exp\{\bm{\gamma}^{T}\bm{X}_{i}^{D}+P_{k}^{v}\}\big\}}{\sum_{m=1}^{K}w_{m}\exp\big\{\sum_{j=1}^{n_{i}}\big(\delta_{ij}^{R}P_{m}^{u}-H_{0}^{R}(t_{ij}^{R})\exp\{\bm{\beta}^{T}\bm{X}_{ij}^{R}+P_{m}^{u}\}\big) + \delta_{i}^{D}P_{m}^{v}-H_{0}^{D}(t_{i}^{D})\exp\{\bm{\gamma}^{T}\bm{X}_{i}^{D}+P_{m}^{v}\}\big\}}.
\end{equation}
It is worth to notice that $\mathbb{Z}_{ik}$ represents the probability that subject $i$ belongs to point $k$, given the current state of parameters. This allows to identify a latent partition of subjects into the $K$ points.\\

\textbf{M-step:} The Maximization step consists of maximizing $Q\left(\bm{\Omega}\right)$ with respect to $\bm{\Omega}$, given the $\mathbb{Z}_{ik}$ obtained at the E-step. It is useful to notice that the three terms involved in the log-likelihood (\ref{eq:logl}) depend on three disjoint subsets  of parameters: $\bm{\Omega}_{w}=\left[\bm{w}\right]$, $\bm{\Omega}_{R}=\left[\bm{\beta}, H_0^R(t), \bm{\mathcal{P}}_u\right]$, and $\bm{\Omega}_{D}=\left[\bm{\gamma}, H_0^D(t), \bm{\mathcal{P}}_v\right]$, where $\bm{\mathcal{P}}_u$ and $\bm{\mathcal{P}}_v$ are respectively the vectors of abscissas and ordinates of the points composing the support of the discrete distribution. The maximization of $Q\left(\bm{\Omega}\right)$ can be carried out separately with respect to these three terms: $Q_w\left(\bm{\Omega}_w\right) := \EX_{\bm{\mathcal{Z}}|\hat{\bm{\Omega}}}\left[ \ell_{w}\left(\bm{\Omega}_{w};\bm{\Theta}\right)\right]$, 
$Q_R\left(\bm{\Omega}_R\right) :=
\EX_{\bm{\mathcal{Z}}|\hat{\bm{\Omega}}}\left[ \ell_{R}\left(\bm{\Omega}_{R};\bm{\Theta}_R\right)\right]$, and 
$Q_D\left(\bm{\Omega}_D\right) :=
\EX_{\bm{\mathcal{Z}}|\hat{\bm{\Omega}}}\left[ \ell_{D}\left(\bm{\Omega}_{D};\bm{\Theta}_D\right)\right]$.

Recalling that weights $w_k$ must sum up to 1, the maximization of $Q_w\left(\bm{\Omega}_w\right)$ is a constrained optimization problem. Using Lagrangian optimization we obtain
\begin{equation}\label{eq:Mstep:w}
    \hat{w}_{k}=\frac{1}{N}\sum_{i=1}^{N}\mathbb{Z}_{ik} \ \ \ \forall k=1,...,K.
\end{equation}

The optimization of $Q_R\left(\bm{\Omega}_R\right)$ involves multiple parameters, so we adopt a multi-step approach. First, we estimate the abscissas $\bm{\mathcal{P}}_u$ of the support points, fixing $\bm{\beta}$ and ${H}_{0}^{R}(t)$ to their last available estimates:
\begin{equation}\label{eq:Mstep:Pu}
    \hat{P}_{k}^{u}=\log \left[ \frac{\sum_{i=1}^{N}\mathbb{Z}_{ik}\sum_{j=1}^{n_{i}}\delta_{ij}^{R}}{\sum_{i=1}^{N}\mathbb{Z}_{ik}\sum_{j=1}^{n_{i}}{H}_{0}^{R}(t_{ij}^{R})\exp\left(\bm{\beta}^{T}\bm{X}_{ij}^{R}\right)}\right] \ \ \ \forall k=1,...,K.
\end{equation}
By substituting $\hat{\bm{\mathcal{P}}}_u$ in $Q_R$ and recalling that $\sum_{k=1}^{K} \mathbb{Z}_{ik} = 1$, we can rewrite $Q_R$ in the following form:
\begin{align}
 \label{eq:QR2}
     Q_{R}\left(\bm{\beta},H_{0}^{R}(t)\big|\hat{\bm{\mathcal{P}}_u}\right)=\sum_{i=1}^{N}\sum_{j=1}^{n_{i}}\Bigg[ & \delta_{ij}^{R}\cdot \left\{\log\left(h_{0}^{R}\left(t_{ij}^{R}\right)\right)+\bm{\beta}^{T}\bm{X}_{ij}^{R}+\sum_{k=1}^{K}\mathbb{Z}_{ik}\hat{P}_{k}^{u} \right\} +\\&  -H_{0}^{R}\left(t_{ij}^{R}\right) \cdot \left\{\sum_{k=1}^{K}\mathbb{Z}_{ik}\cdot\exp\left(\hat{P}_{k}^{u} \right)  \right\}\cdot\exp\left(\bm{\beta}^{T}\bm{X}_{ij}^{R}\right)\Bigg] \nonumber
 \end{align}
that is the usual full log-likelihood of a Cox model with known offset $\log \left[\sum_{k=1}^{K}\mathbb{Z}_{ik}\cdot\exp\left(\hat{P}_{k}^{u} \right)  \right]$. With arguments similar to \cite{johansen1983} and \cite{Gasperoni}, we can then compute the Breslow estimator for the cumulative baseline hazard for recurrent events as follows:
\begin{equation}\label{eq:Mstep:HR}
    \hat{H}_{0}^{R}(t)=\sum_{ab:t_{ab}^{R}\le t} \frac{d^R_{ab}}{\sum_{rs \in \mathcal{R}(t^R_{ab})} \left\{\sum_{k=1}^{K}\mathbb{Z}_{rk}\cdot\exp\left(\hat{P}_{k}^{u} \right)  \right\} \cdot \exp\left(\hat{\bm{\beta}}^{T}\bm{X}_{rs}^{R}\right)}
\end{equation}
where $t^R_{ab}$ is the time of recurrent event $b$ for patient $a$, $d^R_{ab}$ is the total number of recurrent events happening at time $t^R_{ab}$ and $\mathcal{R}(t^R_{ab})$ represents the recurrent-risk set at time $t^R_{ab}$. \\
Lastly, including $\hat{H}_{0}^{R}(t)$ in Eq. \eqref{eq:QR2}, we obtain the profile log-likelihood as a function of $\bm\beta$:
\begin{equation}\label{eq:Mstep:profR}
    \ell^{R}_{\text{profile}}(\bm{\beta})=\sum_{i=1}^{N}\sum_{j=1}^{n_{i}}\delta_{ij}^{R}\cdot\left[\bm{\beta}^{T}\bm{X}_{ij}^{R}-d_{ij}^{R} \cdot\log \sum_{ab \in \mathcal{R}(t_{ij}^{R})} \left\{\mathbb{Z}_{ak}\cdot\exp\left(\hat{P}_{k}^{u} \right)  \right\} \cdot
    \exp\left(\bm{\beta}^{T}\bm{X}_{ab}^{R}\right)\right]
\end{equation}
which is of the form of the usual partial log-likelihood in the Cox model with known offsets, so it is maximized through standard software in order to retrieve $\hat{\bm{\beta}}$.

The optimization of $Q_D\left(\bm{\Omega}_D\right)$ can be performed following the same procedure designed for $Q_R\left(\bm{\Omega}_R\right)$. Similarly, the estimates  $\hat{\bm{\mathcal{P}}}_{v}$ and $\hat{H}_{0}^{D}(t)$ are given by:
\begin{equation}\label{eq:Mstep:Pv}
    \hat{P}_{k}^{v}=\log \left[ \frac{\sum_{i=1}^{N}\mathbb{Z}_{ik}\cdot\delta_{i}^{D}}{\sum_{i=1}^{N}\mathbb{Z}_{ik} \cdot {H}_{0}^{D}(t_{i}^{D})\exp\left(\bm{\gamma}^{T}\bm{X}_{i}^{D}\right)}\right] \ \ \ \forall k=1,...,K;
\end{equation}
\begin{equation}\label{eq:Mstep:HD}
    \hat{H}_{0}^{D}(t)=\sum_{a:t_{a}^{D}\le t} \frac{d^R_{a}}{\sum_{r \in \mathcal{R}(t^D_{a})} \left\{\sum_{k=1}^{K}\mathbb{Z}_{rk}\cdot\exp\left(\hat{P}_{k}^{v} \right)  \right\} \cdot \exp\left(\hat{\bm{\gamma}}^{T}\bm{X}_{r}^{D}\right)}
\end{equation}
where $t^D_{a}$ is the terminal event time for patient $a$,
$d^D_{a}$ is the number of terminal events happened at $t^D_{a}$ and $\mathcal{R}(t^D_{a})$ is the terminal-risk set at time $t^D_{a}$. Regression parameters $\hat{\bm{\gamma}}$ can be retrieved by maximizing the following partial profile log-likelihood:
\begin{equation}\label{eq:Mstep:profD}
    \ell^{D}_{\text{profile}}(\bm{\gamma})=\sum_{i=1}^{N} \delta_{i}^{D}\cdot\left[\bm{\gamma}^{T}\bm{X}_{i}^{D}-d_{i}^{D} \cdot\log \sum_{a \in \mathcal{R}(t_{i}^{D})} \left\{\mathbb{Z}_{ak}\cdot\exp\left(\hat{P}_{k}^{v} \right)  \right\} \cdot
    \exp\left(\bm{\gamma}^{T}\bm{X}_{a}^{D}\right)\right].
\end{equation}

\subsubsection{Computation of standard errors.}
\label{sec:est_NDStdErrors}
To estimate the standard errors of the coefficients, we used the conventional \texttt{coxph} approach of Cox models from \texttt{survival} package \citep{survivalpack}, where the inverse of the Hessian matrix is evaluated at the estimated coefficients. This enables us to assess the statistical significance of the parameters by calculating the corresponding Wald statistic \citep{waldtest}. Similarly, we apply the same approach to obtain standard errors for the discrete random effects. $\forall k=1,...,K$, we compute the Hessian $\mathcal{H}$ as the second derivative of $Q_{R}(\mathbf{\Omega}_R)$ with respect to $P_{k}^u$ and we evaluate it at $\hat{P}_{k}^u$: 
\begin{align}
&\EX \left[\mathcal{H}(\hat{P}_{k}^u)\right] = \frac{\partial^2 \mathbf{\Omega}_R}{\partial {{{\mathcal{P}}}}_{u} \partial {{{\mathcal{P}}}}_{u}'} (\hat{P}_{k}^u)=
-\sum_{i=1}^{N}\mathbb{Z}_{ik}\sum_{j=1}^{n_{i}}{H}_{0}^{R}(t_{ij}^{R})\exp\left(\bm{\beta}^{T}\bm{X}_{ij}^{R}\right) \exp\left(\hat{P}_{k}^u \right)\\
&Var[\hat{P}_{k}^u] = \left(\mathcal{I}(\hat{P}_{k}^u)\right)^{-1} = \left(-\EX \left[\mathcal{H}(\hat{P}_{k}^u)\right]\right)^{-1}
\end{align}

\noindent
where $\mathcal{I}$ represents the information matrix. Equivalently, we compute the variance of $\hat{P}_{k}^v$.

\subsection{Support points identification}
\label{sec:est_supportReduction}
Up to now, we have detailed the steps of the EM algorithm by considering the total number  $K$ of support points of the discrete distribution as known. In order to estimate the discrete distribution on $\mathbb{R}^{2}$, we propose a wrapper method that, given an initial grid, performs a support reduction, according to \cite{Masci1}.\\

\textbf{Grid Initialization:}
The initial step involves defining a grid of $M$ points in $\mathbb{R}^{2}$ that ideally covers the region believed to contain the true support of the discrete distribution (which is unknown). This can be accomplished based on existing knowledge, such as insights from general exploratory analysis, medical expertise, or previously fitted models. 
Another option could be sampling from a specific distribution, such as:
\begin{itemize}
\item[(i)] a \textit{bivariate Gaussian} distribution, where weights are initialized according to the corresponding Normal density and then normalized to be unitary \citep{recurrent6};
\item[(ii)] a \textit{Uniform} distribution of points over a rectangle in $\mathbb{R}^{2}$, whose boundaries are defined to cover the supposed area of the true support.
\end{itemize}
To account for the algorithm's sensitivity to grid initialization, it is advisable to employ a general and non-informative initialization strategy. This helps mitigate potential misspecification issues. Additionally, in both cases, it is important to ensure that the number $M$ of points in the initial grid is sufficient to adequately explore the designated region.\\

\textbf{Support Reduction:}
Given the initial grid, the EM algorithm (Section \ref{sec:est_EM}) gradually performs the support reduction of the discrete distribution to identify $K<M$ mass points. At each E-M iteration, this reduction involves two steps.
\begin{enumerate}
    \item[(I)] First, prior to the E-step, a specified threshold $L$ is defined and the merging process is performed: if two points $\bm{P}_{m_1}$ and $\bm{P}_{m_2}$ are closer than $L$, in terms of a pre-defined distance metric, they collapse at a unique point $\bm{P}_{\bar{m}}=\left(\frac{P_{m_1}^u+P_{m_2}^u}{2}, \frac{P_{m_1}^v+P_{m_2}^v}{2}\right)$ with weight $w_{\bar{m}} = w_{m_1} + w_{m_2}$. 
    The merging process begins with the pair having the minimum distance less than $L$ and continues until no remaining pairs closer than the threshold remain, resulting in $\tilde{K}$ mass points. 
    This process is sensitive to two design choices: the threshold $L$ and the distance metric used for merging points. Specifically, the Euclidean distance offers a clear interpretation of the merging criterion based on geometric distance. However, the Manhattan distance could be beneficial when dealing with specific patterns in the hidden discrete distribution of random effects. Further insights into the definition of threshold $L$ are discussed in Section \ref{sec:est_threshold}.
    Subsequently, the E-step is executed and the $\mathbb{Z}_{ik}^{(I)}$ are computed for each individual $i$ and remaining support point $k=1,...,\tilde{K}$. 

    \item[(II)] Prior to proceeding to the M-step, each individual $i$ is assigned to the sub-population (i.e., mass point) $k_i^*$ such that $ k_i^* = \text{argmax}_{k} \mathbb{Z}_{ik}$ and the support points that do not contain any individual are deleted. 
    Let $\mathcal{K}^{*}$ denote the set of remaining mass-points with $|\mathcal{K}^{*}|=K$. When one or more mass points are deleted (i.e., $\tilde{K} \neq K$), the probabilities that individual $i$ belongs to mass-points $k \in \mathcal{K}^{*}$ are re-parameterized in such a way that they sum up to 1:
    \begin{equation}
    \mathbb{Z}_{ik}^{\text{new}} = \frac{\mathbb{Z}_{ik}^{\text{old}}}{\sum_{m\in \mathcal{K}^{*}}^{} \mathbb{Z}_{im}^{\text{old}}}.
    \end{equation}
    Finally, the M-step is executed by considering in Eq. \eqref{eq:Mstep:w} the remaining re-parametrized probabilities $\mathbb{Z}_{ik} = \mathbb{Z}_{ik}^{\text{new}}$ with $k=1,\dots, K$.
\end{enumerate}

The algorithm terminates when the number of masses in the discrete distribution is stable (i.e., no reduction happens in the current iteration) and the maximum difference between the components of the weights of the current and previous iteration is less than a stopping threshold $toll$, or when a predefined number of iterations $max.it$ is reached. The overall procedure is summarized in panel Algorithm \ref{alg:GridShrink}.

\begin{algorithm}[h]
    \caption{Estimation procedure of JMDF}\label{alg:GridShrink}
    \noindent \textbf{Input parameters:}\\
    $M$: initial number of support points\\
    $init$: type of initialization procedure (Gaussian or Uniform) with relative initial parameters\\
    $distance$: type of distance (Euclidean, Manhattan,...)\\
    $L$: value of threshold for the merging process\\
    $max.it$: maximum number of iterations\\
    $toll$: stopping threshold\\
    \textbf{Estimation procedure:}
    \begin{algorithmic}[1] 
    \State \textbf{\textit{Grid initialization}} $\left[\bm{\mathcal{P}},\bm{w}\right]^{(0)}$ according to the $init$ procedure with (at least) $M$ support points 
    \State \textbf{\textit{Parameter initialization}} $\left[\bm{\beta},\bm{\gamma},H_0^R(t),H_0^D(t)\right]^{(0)}$ by distinct Cox-type models with offsets $\bm{\mathcal{P}}_u^{(0)}$ and $\bm{\mathcal{P}}_v^{(0)}$
    \State Set iteration $it=0$, $converged = FALSE$, and $K^{(0)}=M$
   \While{$!converged \And it\le max.it$}
    \State Update iteration: $it = it + 1$
    \State{\textbf{\textit{Support Reduction I}}: merge points closer than $L$ (in terms of $distance$) by averaging the components and adding up their weights}
    \State{\textbf{\textit{E-step}}: compute $\mathbb{Z}_{ik}^{(it)}$ for each patient $i$ and remaining support point $k=1,\dots,\tilde{K}^{(it)}$}
    \State{\textbf{\textit{Support Reduction II}}: extract the latent partition, delete empty support points, and re-parameterise the remaining conditional probabilities $\mathbb{Z}_{ik}^{(it),\text{new}}$ with $k=1,\dots,K^{(it)}$}
    \State{\textbf{\textit{M-step}}}: update
	$\bm{\Omega}_{w}^{(it)}=\left[\bm{w}\right]^{(it)}$, $\bm{\Omega}_{R}^{(it)}=\left[\bm{\beta}, H_0^R(t), \bm{\mathcal{P}}_u\right]^{(it)}$, and $\bm{\Omega}_{D}^{(it)}=\left[\bm{\gamma}, H_0^D(t), \bm{\mathcal{P}}_v\right]^{(it)}$
	\If{$K^{(it)}=K^{(it-1)} \And \text{max}_k \big|w_k^{(it)}-w_k^{(it-1)}\big|<toll$}
	\State {$converged=TRUE$}
	\EndIf
	\EndWhile
	\State Resulting estimates for parameters $\bm{\Omega}$ and number of mass points $K$ are $ \left[\bm{\mathcal{P}},\bm{w},\bm{\beta},\bm{\gamma},H_0^R(t),H_0^D(t)\right]^{(it)}$ and $K^{(it)}$, respectively.
    \end{algorithmic}
\end{algorithm}

\subsubsection{Definition of threshold $L$}
\label{sec:est_threshold}
Defining the threshold $L$ that determines which points will be collapsed is a crucial aspect of the estimation procedure, as it has a significant impact on the resulting discrete distribution and the identified number of masses. In general, it is advisable to set the threshold to the smallest value that captures a meaningful difference in subject classification for the specific application, taking into consideration available knowledge.
From a practical standpoint, conducting a sensitivity analysis is recommended. This involves examining the behavior according to a fitting criterion (e.g., log-likelihood, AIC, classification log-likelihood) for different threshold values (and different runs for the grid initialization) to identify the most promising candidates. The choice of the fitting criterion itself will influence the sensitivity analysis. 
In this study, to fairly compare models with different number of masses  $K$, we consider the classification likelihood 
\begin{equation}
\mathcal{L}_{\text{class} }\left(\bm{\Omega};\bm{\Theta}|\bm{\mathcal{Z}}\right)=\prod_{k=1}^{K}\prod_{i=1}^{N} \big[\mathcal{L}_{ik}(\bm{\Omega};\bm{\mathcal{O}}_{i}|\bm{z}_i)\big]^{z_{ik}}
\end{equation}
rather than the mixture one \citep{mclachlan19829}, and we computed the AIC and the BIC accordingly, by considering the number of parameters $g = p_1 + p_2 + K\times 2 + (K-1)$.

%%%%% APPLICATION %%%%%%%%%%%%%%%%%%%%%%%%%%%%%%%%%%%%%%%%%%%%%%%%%%
\section{Application}
\label{sec:app}

The approach developed and presented in Section \ref{sec:met_our} is motivated by a study of patients with Heart Failure (HF) undergoing ACE inhibitors treatment, where recurrent events of interest are hospitalizations due to HF and terminal event is death for any cause. In Sections \ref{sec:app_data} and \ref{sec:app_setting}, we introduce the real administrative HF database of \textit{Regione Lombardia - Healthcare Division}  \citep{RegioneLombardia2012} and we present the joint model setting. Results of our method and the comparison with \textit{joint parametric frailty models} by \cite{RondeauJFM} and \cite{NgJFMBiostat} are reported in Section \ref{sec:app_results} and \ref{sec:app_compare}, respectively.

\subsection{Data}\label{sec:app_data}
Administrative data were provided by \textit{Regione Lombardia - Healthcare Division}  within the research project \textit{HFData} [HFData—RF-2009-1483329] \citep{RegioneLombardia2012}. The project database was built for non-paediatric residents in Lombardy (a region in northern Italy)  which were hospitalized for HF from 2000 to 2012. A 5-years period from 2000 to 2005 was used in order to consider only “incident'' HF patients, i.e., patients with no contacts with healthcare system in the previous five years due to HF. 
Each record in the dataset was related to (i) patient ordinary admission to hospital (Hospital Discharge Charts, HDC) -- which contain data related to discharge date, length of stay in hospital and comorbidity conditions assessed as in \cite{Gagne2011}, or (ii) pharmaceutical purchases (identified by their Anatomical Therapeutic Chemical (ATC) codes; see \citealp{WHO2003:dur}) -- which provide information on the number and times of drug purchases. Deaths were collected from the HDC Database (for in-hospital deaths) or Vital Statistics Regional Dataset (for out-hospital deaths). 
Further details regarding data extraction and selection are discussed in \cite{Mazzali2016}.

In this work, we focused on a representative sample of \textit{HFData} related to 4,872 patients with their first HF discharge between January 2006 to December 2012. Overall survival was measured from the index hospitalization to the date of death or to the administrative censoring date (December 31$^{\text{st}}$, 2012). To assess the effect of ACE treatment on both survival and re-hospitalizations, only subjects who experienced at least one hospitalization and one ACE purchase after the index event were selected. Demographics, comorbidities and adherence to ACE drugs were considered to adjust models.
In particular, a dichotomous time-dependent variable that at each event-time indicates whether the patient was adherent to ACE therapy according to the proportion of days covered method with an 80\% threshold was used \citep{Spreafico2021BJ1}.

\subsubsection{Descriptive statistics.} \label{sec:app_data:descr}
A final cohort of $N=2,970$ patients who underwent ACE inhibitors therapy and experienced at least one re-hospitalization was selected. At index event, median age and number of comorbidities were 74 years (IQR = [67; 80]) and 2 (IQR=[1; 3]) respectively, with a percentage of male patients equal to 58.1\% (1,726 patients). 
Before the terminal event/censoring, median number of total re-hospitalizations was 3 (IQR = [2; 6]), with a maximum of 42 occurrences. At last hospitalization event, median age and number of comorbidities were 77 years (IQR = [70; 83]) and 3 (IQR=[2; 5]) with 1,058 patients (35.6\%) adherent to ACE therapy. Median overall survival and final gap-time (i.e., time between the last recurrence and the terminal event/censoring) computed using the reverse Kaplan-Meier method by \cite{schemper1996} were 1,894 days (IQR=[1,365; 2,265]; about 5.1 years) and 485 (IQR=[197; 1,053]; about 1.3 years), respectively. At death/censoring event, 1,032 patients (34.7\%) resulted adherent to ACE therapy and 2,139 (72.0\%) patients were alive.\\
To proceed with the analyses, administrative data was reformatted as explained in Appendix \ref{appendix:setting}.

\subsection{Joint frailty models for re-hospitalizations and death} \label{sec:app_setting}
In order to assess the role of patient's clinical history on both re-hospitalizations and death through the JMDF in Eq. \eqref{eq:ourJM}, we assumed that the two instantaneous hazards for each patient depend on four explanatory variables: \texttt{sex} (\textit{male} or \textit{female}; time-fixed), \texttt{age} (in years; time-dependent), number of comorbidities (\texttt{ncom}; time-dependent) registered at the last known hospitalization, and  the dichotomous \texttt{adherent} variable (0/1; time-dependent) indicating whether the patient is adherent or not to ACE therapy at the considered event time \citep{Spreafico2021BJ1}. The proposed JMDF for re-hospitalizations ($R$) and death ($D$) via discretely-distributed non-parametric random effects was given by  
\begin{equation}\label{eq:ourJM_appHF}
    \begin{cases}
    h^{R}_{ij}\left(t^{R}_{ij}\right) = h^{R}_{0}\left(t^{R}_{ij}\right)\exp\left( \beta_1 \texttt{sex}_i +  \beta_2 \texttt{age}_{ij} +  \beta_3 \texttt{ncom}_{ij} + \beta_4 \texttt{adherent}_{ij} + u_{i}
    \right)\\
    h^{D}_{i}\left(t^{D}_{i}\right) =h^{D}_{0}\left(t^{D}_{i}\right)\exp\left( \gamma_1 \texttt{sex}_i +  \gamma_2 \texttt{age}_{in_i} +  \gamma_3 \texttt{ncom}_{in_i} + \gamma_4 \texttt{adherent}_{in_i} + v_{i}
    \right) 
    \end{cases}
\end{equation}
where $h_{0}^{R}(\cdot)$ and $h_{0}^{D}(\cdot)$ are the baseline hazard functions of the re-hospitalizations and death,  respectively, and $[u,v]_{i}$ are the random effects of the $i$-th patient distributed according to $P^{*}$ as in Eq. \eqref{eq:Pstar}. The vectors of parameters $\bm{\beta}=\left(\beta_1,\beta_2,\beta_3,\beta_4\right)$ and  $\bm{\gamma}=\left(\gamma_1,\gamma_2,\gamma_3,\gamma_4\right)$ are respectively relative to the (time-dependent) vectors of covariates $\bm{x}_{ij}^{R} = \left(\texttt{sex}_i, \texttt{age}_{ij}, \texttt{ncom}_{ij}, \texttt{adherent}_{ij} \right)$ and $\bm{x}_{i}^{D} = \left(\texttt{sex}_i, \texttt{age}_{in_i}, \texttt{ncom}_{in_i}, \texttt{adherent}_{in_i} \right)$. Note that the covariate values in $\bm{x}_{ij}^{R}$ may vary for each event $j$  experienced by a patient (except for sex), whereas the values in $\bm{x}_{i}^{D}$ are taken at the last gap-time $n_i$, representing the patient's last available measurement (see Appendix \ref{appendix:setting}). 

The JMDF applied to \textit{HFData} proposed in Eq. (\ref{eq:ourJM_appHF}) was compared to the joint frailty models introduced by \cite{RondeauJFM} and \cite{NgJFMBiostat} (see Sections \ref{sec:met_sa_rondeau} and \ref{sec:met_sa_ng}, respectively).
The setting was the same as in Eq. (\ref{eq:ourJM_appHF}), except for the random effects structure:
\begin{itemize}
\item \cite{RondeauJFM} assume shared log-Normal random effects with $u_i = \eta_i$ and $v_i =\alpha \eta_i$,
where $\eta_i \sim \mathcal{N}(0, \sigma^2)$ is the patient-specific random intercept in the recurrent events process and $\alpha$ is the multiplicative parameter which quantifies the effect of the patient frailty on the terminal event process;
\item \cite{NgJFMBiostat} assume bivariate Gaussian random effects as in Eq. \eqref{eq:Ngform}.
\end{itemize}

\subsection{Results of the joint models with discrete non-parametric frailty}\label{sec:app_results}
The results of applying the JMDF in Eq. \eqref{eq:ourJM_appHF} to the cohort presented in Section \ref{sec:app_data:descr} are now discussed step-by-step.

\subsubsection{Grid initialization and identification of threshold $L$}
For each initialization procedure (Gaussian and Uniform), we conducted multiple runs of the algorithm while varying the threshold values $L$ (measured in terms of Euclidean distance) from 0.1 to 3 to determine the optimal one.

\begin{figure}[ht]
    \centering \includegraphics[width=0.9\textwidth]{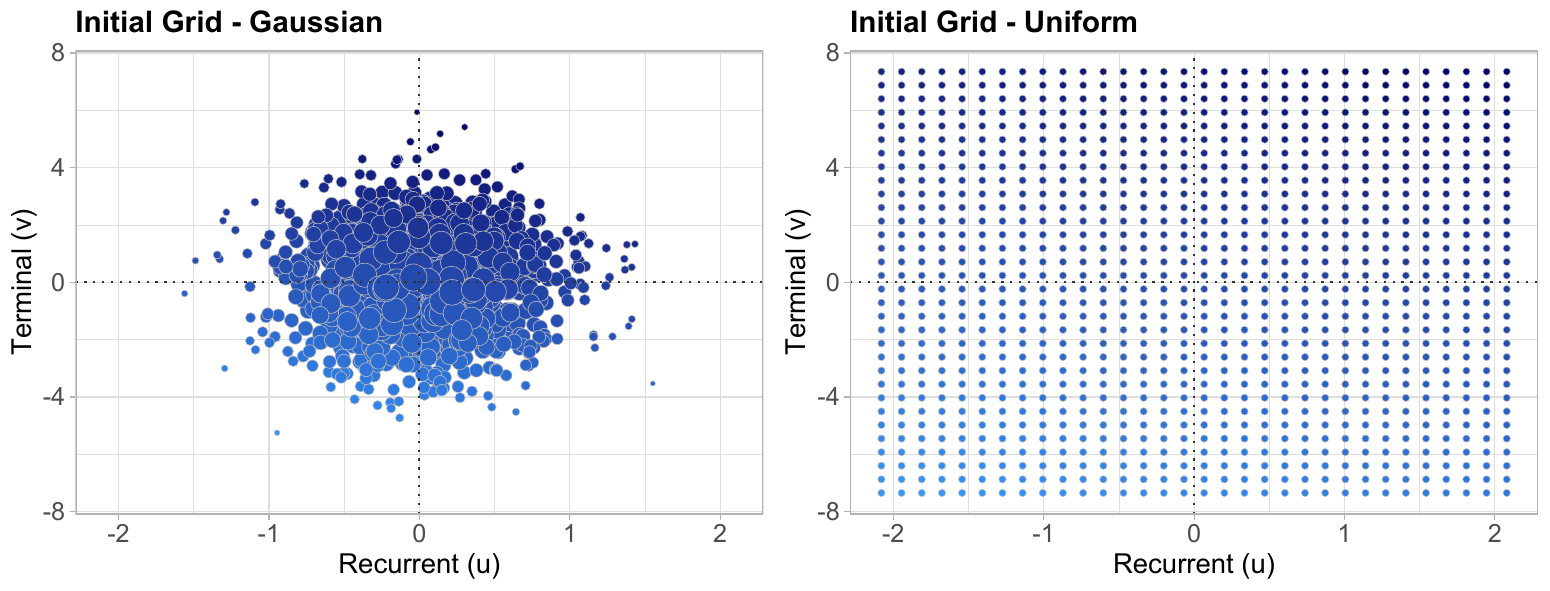}
    \caption{Gaussian (left panel) and uniform (right panel) initial grids for the random effects distribution of the discrete frailty model. Each point is colored according to a gradient scale from blue to red, which distinguishes
    points associated with decreased risk (blue) from one associated with increased risk (red). The size of each point reflects
    its weight in the discrete distribution.}
    \label{fig:grids}
\end{figure}
\begin{figure}[h!]
    \centering \includegraphics[width=1\textwidth]{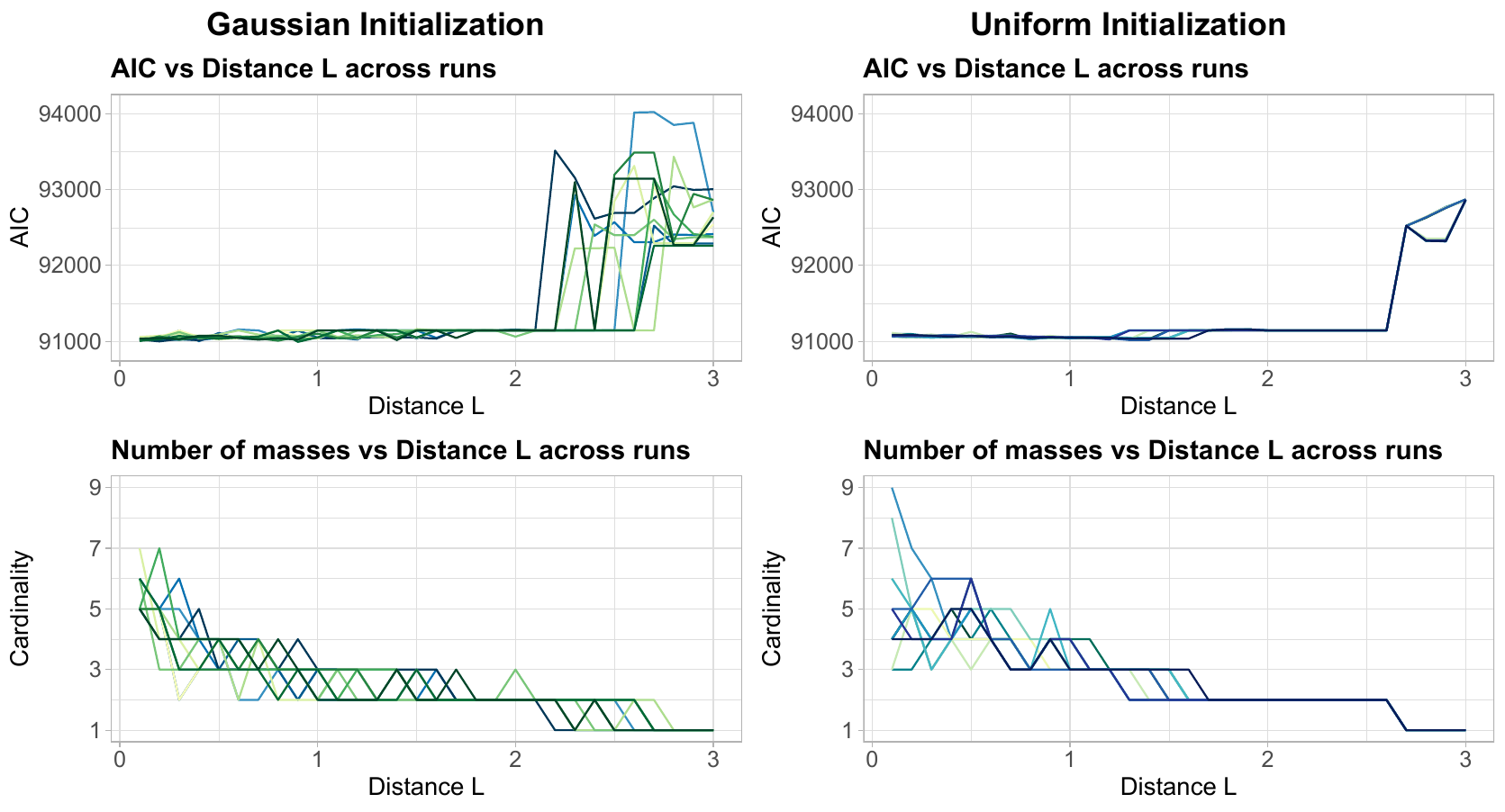}
    \caption{AIC (top panels) and number of masses (bottom panels) obtained by the joint model with discrete frailty under different distance thresholds $L$ and across 12 runs of the algorithm. Left and right panels refer to the  Gaussian and Uniform initialization cases, respectively.}
    \label{fig:sens_L}
\end{figure}

At each run with Gaussian initialization, we sampled $M=1000$ points from a bivariate Gaussian distribution centered at the origin in $\mathbb{R}^{2}$ and with a diagonal variance-covariance matrix. 
To ensure a proper exploration of the space and prevent an overly informed distribution, we opted to double the variance estimated by the joint model with bivariate Gaussian random effect by \cite{NgJFMBiostat} (see Section \ref{sec:app_compare}) and enforce zero correlation in the initial grid definition. Weights were computed using the density of the bivariate Gaussian distribution under consideration, and then normalized to sum to one.
For each run with Uniform initialization, we sampled a Uniform distribution over a rectangle centered in the origin of $\mathbb{R}^{2}$ with sides' lengths set at six times the standard deviation of the corresponding variance parameters obtained by the application of the joint model by \cite{NgJFMBiostat}. The rectangle area was filled with $M=1024$ equally spaced points with uniform weights. 
Figure \ref{fig:grids} illustrates an example of the support-point grids of the random effects for the Gaussian (left panel) and the Uniform (right-panel) initializations.

Figure \ref{fig:sens_L} reports the AIC (top panels) and the number of masses $K$ (bottom panels) of models obtained from 12 runs, each with varying values of $L$, using the two initialization procedures.
Results on AIC showed that the Uniform case (right panel) leads to more stable and better estimates compared to the Gaussian one (right panel). This confirmed the advantage of using a more comprehensive and adaptable initialization grid. Furthermore, transitioning from the highest number of masses (7 for the Gaussian case; 9 for the Uniform case) to 3 or even 2 masses ($L\approx 1$ or $L\approx 2$) resulted in only a slight decrease in AIC. The sensitivity analysis was repeated by computing the BIC, and the results remained consistent.
In terms of mass cardinality, both the procedures highlighted the presence of stability regions for different values of $L$ ranging from $1$ to $2.5$, where the fitted model suggested the presence of either 3 and 2 masses.
Striking the best balance between model accuracy and complexity, these regions guided the choice of optimal threshold values. In both the Gaussian and Uniform cases, the values of $L$ from the runs that led to the identification of $K=2$ or $K=3$ masses (i.e., $L\approx1$ and $L\approx2$, respectively) with the lowest AIC were selected as viable options.

\subsubsection{Estimated mass points} 
\begin{figure}[hb]
    \centering \includegraphics[width=1\textwidth]{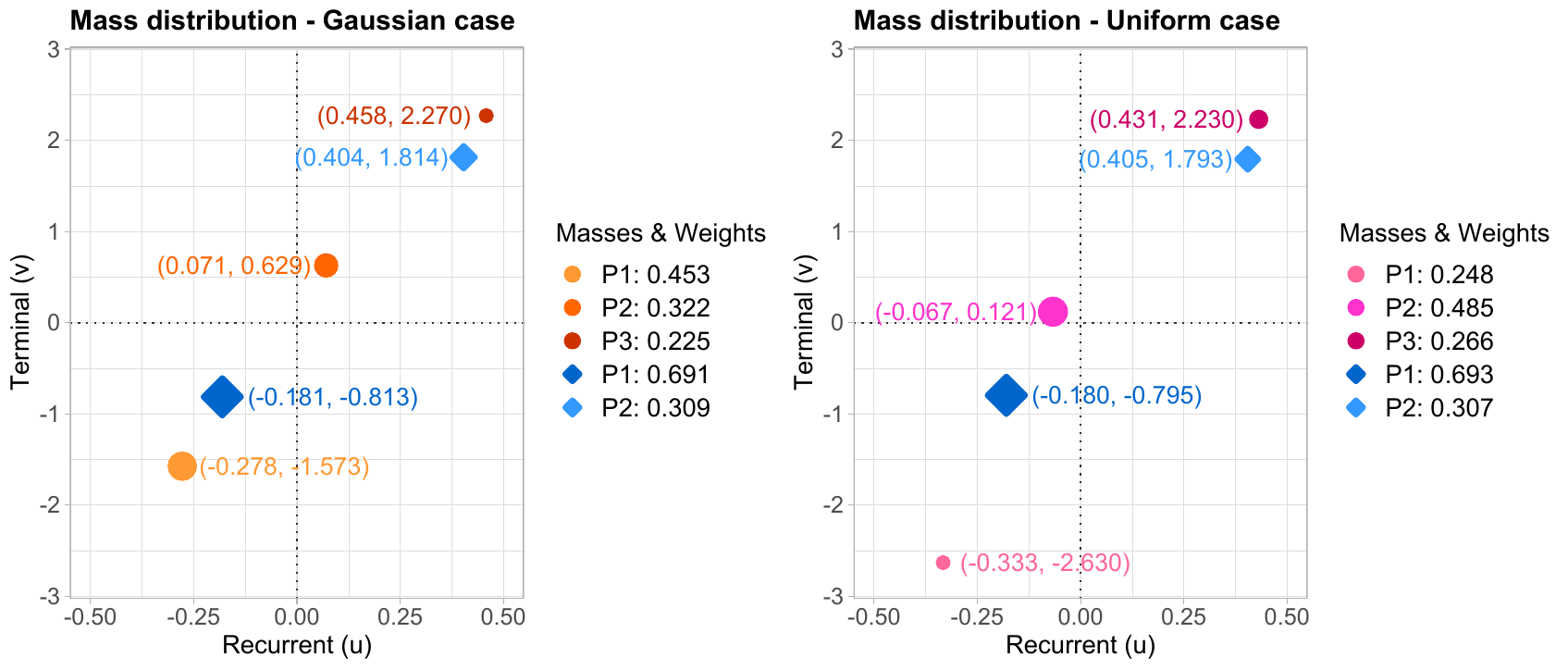}
    \caption{Estimated random-effects of the JMDF specified in Eq. \eqref{eq:ourJM_appHF} with Gaussian (left panel) and Uniform (right panel) initialization procedures, when $K=3$ (circles in orange/pink palette) and $K=2$ (diamonds in blue palette) mass points are identified. Each point is reported along with its coordinates $(u,v)$ and the size is proportional to its mass weight $w$. The complete list of the estimates and their standard errors is reported in Appendix \ref{appendix:df}.}
    \label{fig:randFinal}
\end{figure}

Figure \ref{fig:randFinal} displays the discrete random-effects estimates for the case of 2 and 3 masses, obtained with the Gaussian (left) and Uniform (right) initialization (the complete list of the estimates and their standard errors is reported in Appendix \ref{appendix:df}). The disposition of points follows a linear pattern, suggesting that patients' fragility remains consistent between hospitalizations and death hazards. Notably, the range of values for the estimated random effect associated with the terminal event ($\hat{v}_i$) is broader than that of the recurrent random effect ($\hat{u}_i$).
For both the initialization procedures, when 2 mass points were identified (diamonds in blue palette), the most frequent cluster (about 69\% of patients) was relative to the \textit{neutral/protected} sub-population (blue diamond $\bm{P}_1$) with a slightly negative frailty for both processes. The rest 31\% can be identified as the \textit{at-risk} sub-population (light-blue diamond $\bm{P}_2$), containing individuals at higher risks of both hospital readmission and mortality.
As the value of $L$ decreased to the point where the algorithm identified 3 mass points, we observed that the 2 clusters of patients split into 3 more differentiated clusters. This indicates that the lower value of $L$ enables a finer separation and characterization of the patient population, revealing additional heterogeneity and subgroups within the data.
In this case, results were slightly different for Gaussian and Uniform initialization but in both cases we can distinguish a \textit{protected} sub-population ($\bm{P}_1$), a \textit{neutral} sub-population ($\bm{P}_2$), and the sub-population \textit{at-risk} ($\bm{P}_3$).

This evidence highlights the strength of the proposed methodology in producing straightforward and interpretable results. A comparison of these results with the parametric continuous frailties estimated using the methods by \cite{RondeauJFM} and \cite{NgJFMBiostat} is provided in Section \ref{sec:app_compare}.

\subsubsection{Estimated survival curves stratified by random effects} 
\begin{figure}[h!]
    \centering \includegraphics[width=1\textwidth]{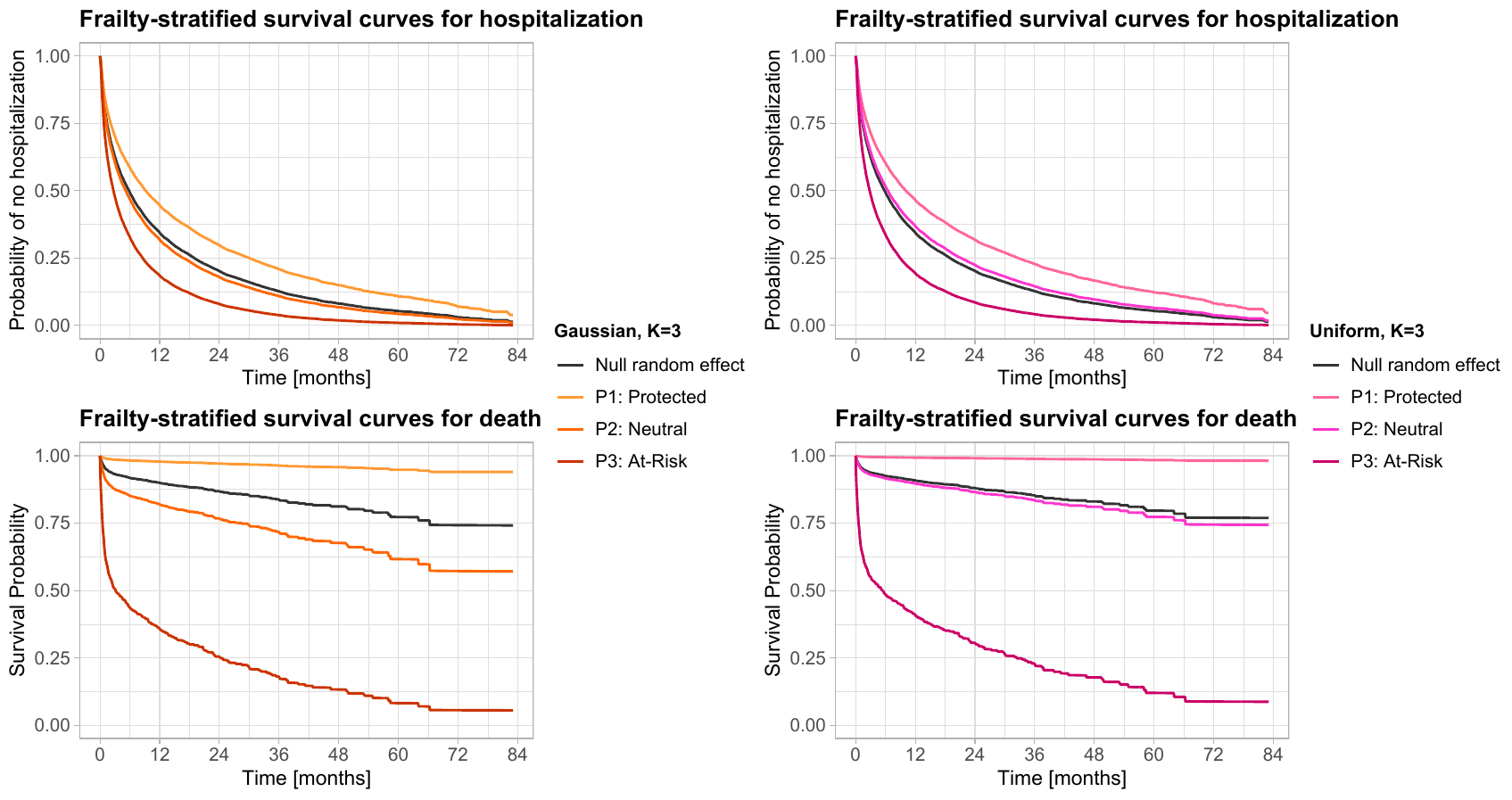}
    \caption{Estimated survival probability curves for hospitalization (top-panels) and death (bottom-panels) processes
    related to a never-adherent male patient aged 74 years at baseline with two comorbidities over time and stratified by random effects. The curves are associated to the discrete frailty distribution identified by using Gaussian (top-left) and Uniform (top-right) initializations when $K=3$ masses are selected. The color of each curve is the same of the corresponding random-effect point as in Figure \ref{fig:randFinal}. The black lines represent the estimated survival probability curves for a null random effect. Time is expressed in days since index event.}
    \label{fig:StratifiedCurves}
\end{figure}

To better quantify and interpret the effects of belonging to different sub-populations, Figure \ref{fig:StratifiedCurves} displays the estimated survival curves for a never-adherent male patient aged 74 years at baseline with two comorbidities over time. The curves are stratified by random effects for both the recurrent and terminal event processes, particularly in the more specific cases of three clusters (left panels: Gaussian; right panels: Uniform).
In each panel, the black line represents the estimated survival curve for a null random effect.
Regarding the re-hospitalization risk (top panels), the distinction among the three sub-populations is subtle yet evident. In both cases, the \textit{at-risk} population exhibits a steeper curve, consistent with their shorter expected time before a new hospitalization.
Aligning with the wider ranges of values for the estimated terminal-process frailties, the difference between the three sub-populations is more pronounced in terms of probability of survival (bottom panels): the \textit{at-risk} sub-population exhibits a high mortality risk since the beginning of the follow-up, in contrast to the \textit{neutral} sub-population, while the \textit{protected} sub-population demonstrates good survival.\\

\subsubsection{Effects of fixed covariates} 
\begin{figure}
    \centering \includegraphics[width=1\textwidth]{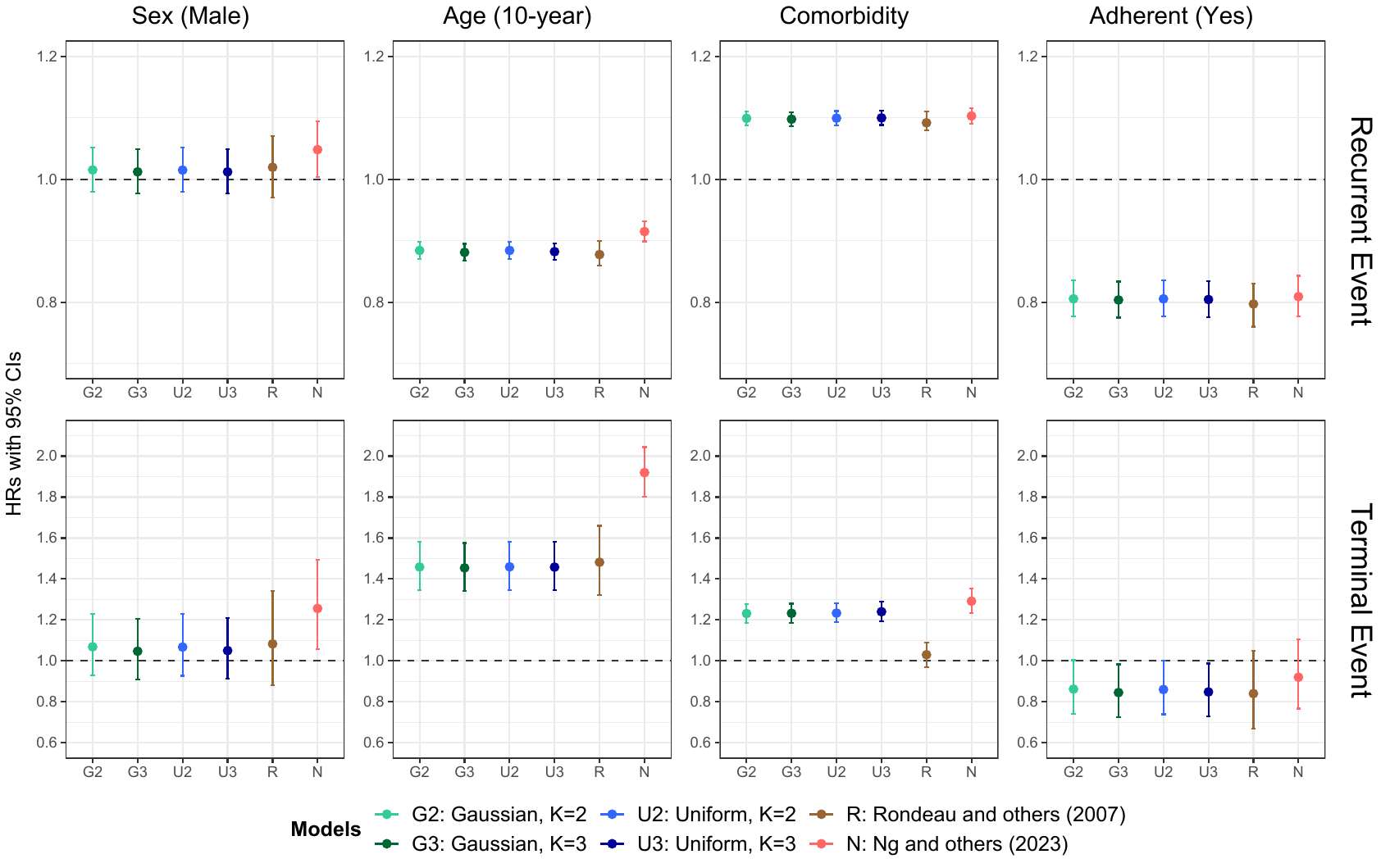}
    \caption{Comparison of estimated fixed-effect Hazard Ratios (HRs) and their 95\% CIs in the trained models. Top panels refer to the recurrent hospitalization process, whereas bottom panels to terminal death process. Considered joint models are: JMDF with Gaussian initialization when $K=2$ (\textit{G2}: green) and $K=3$ (\textit{G3}: dark-green) masses are identified; JMDF with Uniform initialization when $K=2$ (\textit{U2}: light-blue) and $K=3$ (\textit{U3}: blue) masses are identified; Shared log-normal frailty model by \cite{RondeauJFM} (\textit{R}: brown); Multivariate Gaussian frailty model by \cite{NgJFMBiostat} (\textit{N}: salmon).}
    \label{fig:HRcomparison}
\end{figure}

The estimated fixed-effects (covariates \texttt{sex}, \texttt{age}, \texttt{ncom}, and \texttt{adherent}) were strongly consistent across different runs, choices of $L$, and initialization procedures. 
Figure \ref{fig:HRcomparison} reports the Hazard Ratios (HRs) along with the 95\% Confidence Intervals (CIs) for the fixed-effects estimates in both the recurrent (top panels) and terminal (bottom panels) events. Each panel displays the results for the JMDF with 2 or 3 masses, using both Gaussian and Uniform initializations, in addition to the estimates from the parametric frailty models proposed by \cite{RondeauJFM} and \cite{NgJFMBiostat}. The estimated HRs showed a high level of concordance across the models, especially for the recurrent events process. This suggests that, despite varying assumptions about the random effects, the fixed-effects estimates remain consistent across the models. Another positive observation pertains to the standard errors, which are smaller in the JMDF compared to the others.

The \texttt{sex} variable (first columns) did not emerge as a significant predictor for either the hazard of recurrent or terminal events. 
Patient \texttt{age} was found to be statistically significant for both processes. Its effect on the hospitalization hazard is a $12\%$ reduction in the hazard of hospitalization per 10-year increase ($HR=0.88$). Conversely, a 10-year increase results in a 45\% increase in the death hazard (HR=1.45). From a clinical perspective, this phenomenon can be explained by the fact that as patients get older, the risk of experiencing a new hospitalization is partially replaced by the risk of mortality.
The number of comorbidities \texttt{ncom} resulted a statistically significant risk factor for both processes, leading to a 10\% increase in the risk of hospitalization and a substantial 23\% increase in the risk of death per registered comorbidity ($HR=1.10$ and $HR=1.23$, respectively). This confirmed the well-documented role of comorbidities in increasing mortality and hospitalizations among HF patients \citep{como1,como2}. 
Finally, being \texttt{adherent} to the ACE treatment was found to be statistically significant at any level for the recurrent event process, yielding a 20\% decrease in the hazard of a new hospitalization ($HR=0.80$). For the death process, it was significant only in the 3-mass cases, where it led to a 15\% decrease in the death hazard ($HR=0.85$).  From a clinical perspective, these results endorse the efficacy of ACE inhibitors treatment for HF. It demonstrates a significant reduction in the hospitalization rate, consequently lowering the occurrence of critical HF events in adherent patients throughout their clinical journey, while also enhancing their survival probability.

\subsection{Comparison of random effects with joint parametric frailty models} \label{sec:app_compare}
To compare our proposed method with discrete non-parametric random effects to its counterparts, we applied the joint parametric frailty models presented by \cite{RondeauJFM} and \cite{NgJFMBiostat} to the \textit{HFData}.

\begin{figure}[h]
    \centering \includegraphics[width=1\textwidth]{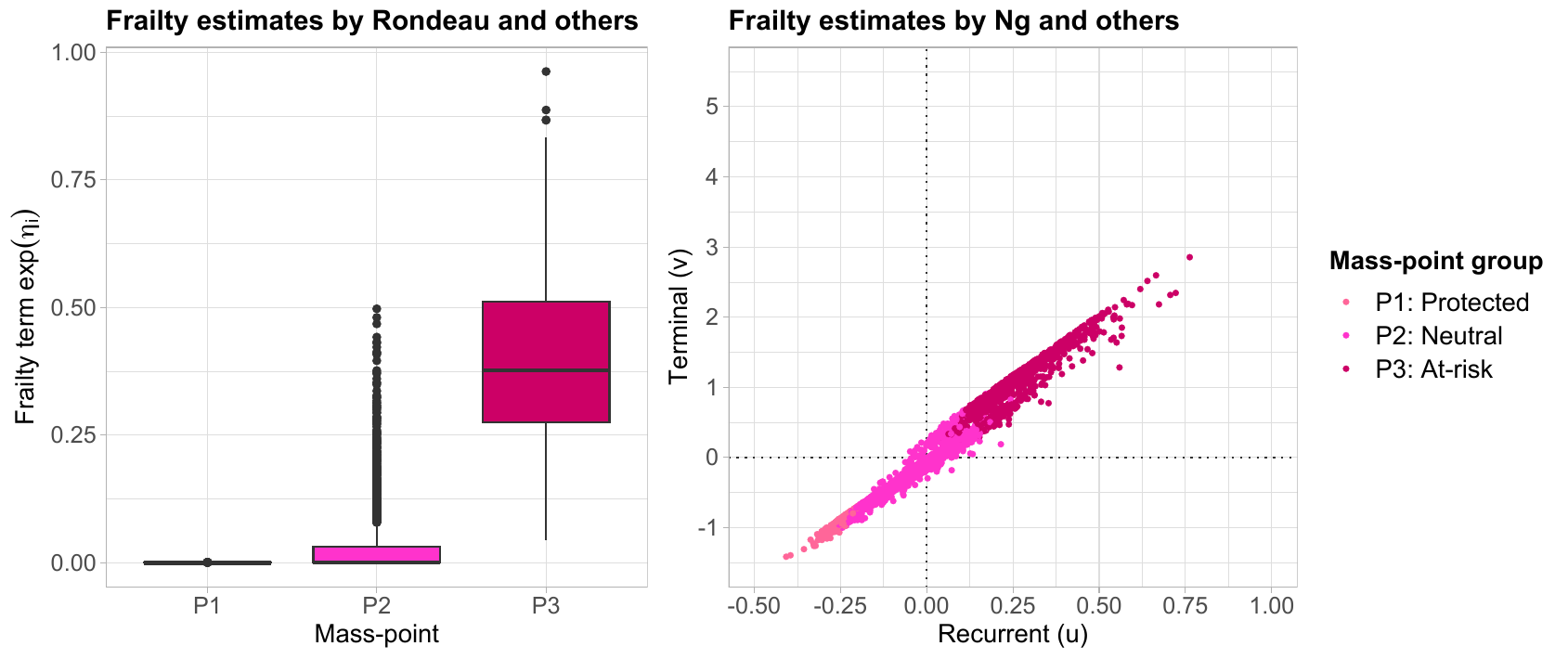}
    \caption{Random effects pointwise estimates for models by \cite{RondeauJFM} (left panel) and \cite{NgJFMBiostat} (right panel). In the right panel, points are visualized in $\mathbb{R}^{2}$ considering hospitalization frailties as abscissa and death frailties as ordinata. Each point correspond to a different subject and it is colored according to the assigned discrete point-mass group for Uniform initialization when $K=3$ masses are identified as in Figure \ref{fig:randFinal}.In the left panel, the distribution of $\exp(\eta_i)$ is stratified according to the assigned 3 discrete mass points.}
    \label{fig:ngREpostprocessing}
\end{figure}

Figure \ref{fig:ngREpostprocessing} displays
pointwise estimates for models by \cite{RondeauJFM} (left panel) and \cite{NgJFMBiostat} (right panel). 
In the right panel, each point corresponds to a different subject $i$ with recurrent $\hat{u}_i$ as abscissa and terminal $\hat{v}_i$ as ordinata, and it is colored according to the assigned discrete point-mass group for Uniform initialization when $K=3$ masses are identified as in Figure \ref{fig:randFinal}. In the left panel, we report the estimated $\exp(\eta_i)$ stratified according to the assigned 3 discrete mass points.
Results are in line with the role of the \textit{protected}/\textit{neutral}/\textit{at-risk} sub-populations.

Table \ref{tab:re_comp} reports the estimated parameters for the random-effects in Figure \ref{fig:ngREpostprocessing}. 
In both models, the frailty associated with the recurrent events process exhibits lower variability compared to the frailty associated with the death process. This observation aligns with the estimated distribution of our mass points (see Figure \ref{fig:randFinal}) and is clinically  reasonable, as subjective factors influencing mortality outcomes may exhibit greater diversity and significance compared to those affecting hospitalizations.
The notably high positive value estimated for the multiplicative parameter in \cite{RondeauJFM} ($\hat\alpha=5.552$) and the strong positive correlation ($\hat{\rho}=0.883$) between the frailties estimated by \cite{NgJFMBiostat} indicate a significant positive correlation between the frailties associated with the two processes, affirming the positions of the mass points in our discrete distribution.  This finding emphasizes the importance of modeling the dependence between the frailties of the recurrent events process and the death process. It highlights that individuals more susceptible to re-hospitalizations are also more likely to experience higher mortality risks. By incorporating and accounting for this dependence in the model, we can better capture the interplay and shared underlying factors between these two processes.

\begin{table}[h]
\centering
\begin{tabular}{ccccc}
\hline
\textbf{Joint model} & \textbf{Random-effect Parameters} & \textbf{Estimate} & \textbf{StdDev} & \textbf{pvalue} \\ \hline
\multirow{2}{*}{\cite{RondeauJFM}} & $\sigma^2$  & 0.127 & 0.009  & $<$2e-16 \\
& $\alpha$   & 5.552 & 0.318 & $<$2e-16 \\ \hline
\multirow{3}{*}{\cite{NgJFMBiostat}} &  $\theta_{u}^2$  & 0.094 & 0.004 & $<$2e-16 \\
& $\theta_{v}^2$     & 1.438 & 0.059 & $<$2e-16 \\ 
& $\rho$  & 0.883 & 0.006 & $<$2e-16 \\ \hline
\end{tabular}%
\caption{Estimated random-effects coefficients of the joint parametric frailty models by \cite{RondeauJFM} and \cite{NgJFMBiostat}.}
\label{tab:re_comp}
\end{table}

%%%%% DISCUSSION %%%%%%%%%%%%%%%%%%%%%%%%%%%%%%%%%%%%%%%%%%%%%%%%%%
\section{Discussion}
\label{sec:discussion}

This paper contributes to the literature on joint models for recurrent and terminal events by introducing an innovative joint frailty model, called JMDF, in which the frailties related to the two processes assume a discrete distribution with an a priori unknown number of support points.
The JMDF approach allows for the clustering of the highest-level units, i.e., patients in our case, based on their frailty levels. After adjusting for the observable factors, the bivariate discrete frailty takes into account the heterogeneity at the patient level, associated with unobserved covariates, and captures the correlation between the two processes. Specifically, the assumption of a discrete distribution translates this heterogeneity into the identification of clusters of patients, enabling us to distinguish different patients profiles based on their associated frailty values. 

The advantage of this modelling approach is twofold. First, using a discrete distribution with an unknown number of support points for frailties can indeed increase the model's flexibility compared to classic parametric distributions. Parametric distributions impose specific assumption on the shape of the frailty that might not always hold in real-world data. By using a discrete distribution, the frailties can encompass a broader range of values and patterns, potentially reducing the risk of bias and leading to a better fit to the data. Second, discrete frailties capture different patterns of vulnerability or risk among patients, with each support point corresponding to a distinct subpopulation of patients sharing a similar frailty profile. These subpopulations represent individuals who share unobserved common characteristics, behaviors, or risk factors affecting their outcomes. This lends interpretability to the subpopulations and provides insights for tailoring interventions to address the specific needs of each group.

When applied to \textit{HFData} for modelling the two correlated processes related to hospitalizations and death, the JMDF suggested the presence of three different patients subpopulations, namely the \textit{protected}, \textit{neutral}, and \textit{at-risk} populations. Compared to the methods proposed in \cite{NgJFMBiostat} and \cite{RondeauJFM}, our fixed-effects estimates aligned with those estimated by the counterparts and, moreover, exhibited smaller standard errors. Regarding random effects, the subpopulations identified by JMDF were consistent with the distribution of continuous frailties estimated by \cite{NgJFMBiostat} and \cite{RondeauJFM}. These findings support the reliability of the proposed method and its results.

Alongside these advantages, our approach has also some limitations and possibilities for future developments. 
Firstly, the estimation procedure is highly sensitive to the choice of the parameter $L$, which tunes the spacing between mass points in the frailty discrete distribution. When users have \textit{a priori} knowledge about the magnitude of differences they aim to observe across patient clusters, the parameter $L$ represents a positive key point. Otherwise, a sensitivity analysis is necessary to address this issue and identify potential values for $L$. By exploiting the clusters identified using different $L$ values, we can identify stability regions and compare their goodness-of-fit indices, achieving a balance between model complexity and the ability to capture meaningful patterns. This procedure also allows for result evaluation at different granularity levels, revealing varying degrees of heterogeneity among the subpopulations. Nonetheless, tuning this parameter might be time-consuming and challenging. Further work will be devoted on developing a significance-based method in which the identified clusters differ in terms of statistical significance \citep{ragni2023clustering}. 
Second, the type of discrete distribution we assume assigns each patient to a cluster that describes the dynamic related to both the death and hospitalization processes. The two processes are assumed to arise from a distribution with the same number of mass points. 
However, this assumption may not always reflect the underlying reality as there might be patients sharing the same type of hospitalization process but not the same type of death process and vice-versa. Patient-level heterogeneity can vary between these processes, and the optimal number of clusters to capture the dynamics within the data could differ as well. Therefore, relaxing this assumption and using a more flexible discrete distribution present a promising avenue for future research in this field. By allowing for varying numbers of clusters or mass points in the frailty distribution for each process, researchers can potentially provide a more accurate and nuanced representation of the underlying phenomena. 

Overall, this work enriches the literature on joint frailty models for recurrent and terminal events by embracing discretely-distributed non-parametric frailties. This new methodology empowers the identification of subgroups of patients united by shared frailty attributes. In collaboration with healthcare professionals, this additional information has the potential to better profile patients and, in turn, improve the refinement of their therapeutic pathways.

%%%%%%%%%%%%%%%%%%%%%%%%%%%%%%%%%%%%%%%%%%%%%%%%%%%%%%%
\vspace{1cm}
\small
\noindent\textbf{Software \& Code.}
Software in the form of R code \citep{Rsoftware}, together with a toy sample input data set and complete documentation is available at \url{http://github.com/mspreafico/JMDF}.\\

\noindent\textbf{Acknowledgments.}
The present research has been supported by MUR, grant Dipartimento di Eccellenza 2023-2027.
The authors wish to thank Riccardo Scaramuzza for the seminal analyses he carried out in his MSc thesis, which represented a starting point for the current dissertation.

%% References
%\bibliographystyle{apalike}
%\bibliography{refs.bib}

%% Appendix
\newpage
\normalsize
\appendix
\section*{Appendix}
\section{Data format}\label{appendix:setting}
After selecting the cohort of patients for analysis and identifying the relevant events in each patient's clinical history (see Section \ref{sec:app_data}), we proceeded to reformat the administrative data to adhere to the required format for the \texttt{coxph} function in the \texttt{survival} R package \citep{survivalpack}. Table \ref{tab:app:data} shows an example of reformatted dataset related to two hypothetical patients $i \in \{A,B\}$. 

The dataset contains 11 rows, which corresponds to the sum of the number of gap times $T_{ij}$ for both patient $A$ and patient $B$, i.e., $n_A=5$ and $n_B=6$ respectively. Patient $A$ experienced four re-hospitalization events ($\delta_{Aj}^R=1$ and $\delta_{Aj}^D=0$ for $j=1,\dots,4$), and was censored at the last follow-up ($\delta_{A5}^D=0$ and $\delta_{A5}^R=0$). Patient $B$ experienced five re-hospitalization events ($\delta_{Bj}^R=1$ and $\delta_{Bj}^D=0$ for $j=1,\dots,5$), and died at the last follow-up ($\delta_{B6}^D=1$ and $\delta_{A6}^R=0$). 

The dataset contains four explanatory variables: patient's $\texttt{sex}_i$ (\textit{male} or \textit{female}; time-fixed), time-dependent $\texttt{age}_{ij}$ (in years) and number of comorbidities ($\texttt{ncom}_{ij}$) registered at the last known hospitalization, and time-dependent dichotomous variable $\texttt{adherent}_{ij}$ indicating whether the patient was adherent to ACE therapy according to the proportion of days covered method with an 80\% threshold \citep{Spreafico2021BJ1}.

The (time-dependent) vectors of covariates $\bm{x}_{ij}^{R} = \left(\texttt{sex}_i, \texttt{age}_{ij}, \texttt{ncom}_{ij}, \texttt{adherent}_{ij} \right)$ and $\bm{x}_{i}^{D} = \left(\texttt{sex}_i, \texttt{age}_{in_i}, \texttt{ncom}_{in_i}, \texttt{adherent}_{in_i} \right)$. Note that the covariate values in $\bm{x}_{ij}^{R}$ may vary for each event $j$  experienced by a patient (except for sex), whereas the values in $\bm{x}_{i}^{D}$ are taken at the last gap-time $n_i$, representing the patient's last available measurement. As an example, considering the second and the last gap times for each patient in Table \ref{tab:app:data}, we have the following values:
$$
\bm{x}_{A2}^{R} = \left(\textit{female},\, 67,\, 5,\, 1\right) \qquad \text{ and } \qquad
\bm{x}_{B2}^{R} = \left(\textit{male},\, 77,\, 2,\, 0\right);
$$
$$
\bm{x}_{A}^{D} = \bm{x}_{A6}^{R} = \left(\textit{female},\, 71,\, 5,\, 0\right) \qquad \text{ and } \qquad
\bm{x}_{B}^{D} = \bm{x}_{B6}^{R} = \left(\textit{male},\, 79,\, 4,\, 1\right).
$$

\vspace{1cm}
\begin{table}[ht]
\centering 
\begin{tabular}{ccccccccc}
  \hline
$i$ & $j$ & $\delta_{ij}^R$ & $\delta_{ij}^D$ & $T_{ij}$ & $\texttt{sex}_i$ & $\texttt{age}_{ij}$ & $\texttt{ncom}_{ij}$ & $\texttt{adherent}_{ij}$ \\ 
  \hline
  A & 1 & 1 & 0 & 49 & \textit{female} & 65 & 5 & 1 \\ 
  A & 2 & 1 & 0 & 901 & \textit{female} & 67 & 5 & 1 \\ 
  A & 3 & 1 & 0 & 391 & \textit{female} & 69 & 5 & 1 \\ 
  A & 4 & 1 & 0 & 10 & \textit{female} & 69 & 5 & 1 \\ 
  A & 5 & 0 & 0 & 801 & \textit{female} & 71 & 5 & 0 \\\hline 
  B & 1 & 1 & 0 & 82 & \textit{male} & 77 & 2 & 0 \\ 
  B & 2 & 1 & 0 & 11 & \textit{male} & 77 & 2 & 0 \\ 
  B & 3 & 1 & 0 & 186 & \textit{male} & 77 & 2 & 1 \\ 
  B & 4 & 1 & 0 & 29 & \textit{male} & 77 & 2 & 1 \\ 
  B & 5 & 1 & 0 & 118 & \textit{male} & 78 & 4 & 1 \\ 
  B & 6 & 0 & 1 & 183 & \textit{male} & 79 & 4 & 1 \\ 
   \hline
\end{tabular}
\caption{Example of reformatted dataset.}\label{tab:app:data}
\end{table}

\section{Discrete frailty estimates}\label{appendix:df}

\begin{table}[H]
\centering
\begin{tabular}{cccccccc} 
 && \multicolumn{3}{c}{\textbf{Gaussian initialization}} & \multicolumn{3}{c}{\textbf{Uniform initialization}} \\ \hline \\[-1.8ex]
& & $\widehat{P}^u_k$ & $\widehat{P}^v_k$ & \multirow{2}{*}{$\widehat{w}_k$} & $\widehat{P}^u_k$ & $\widehat{P}^v_k$ & \multirow{2}{*}{$\widehat{w}_k$}\\ 
 && (sd) & (sd) & & (sd) & (sd) & \\ \hline
\multirow{6}{*}{$K=3$} & \multirow{2}{*}{$\bm{P}_1$} & -0.278  & -1.573 & \multirow{2}{*}{0.453}  &  -0.333 & -2.630 & \multirow{2}{*}{0.248}\\
&& (0.0002) & (0.0199) & & (0.0004) & (0.1131) & \\ \cline{2-8}
& \multirow{2}{*}{$\bm{P}_2$} &  0.071  & 0.629 & \multirow{2}{*}{0.322} &  -0.067 & 0.121 & \multirow{2}{*}{0.485}\\ 
&& (0.0002) & (0.0036) & & (0.0002) & (0.0040) & \\ \cline{2-8}
& \multirow{2}{*}{$\bm{P}_3$} &   0.458 &  2.270 & \multirow{2}{*}{0.225} & 0.431 & 2.230 & \multirow{2}{*}{0.266} \\
&& (0.0002) & (0.0019) & & (0.0002) & (0.0017) & \\ \hline 
\multirow{4}{*}{$K=2$} & \multirow{2}{*}{$\bm{P}_1$} & -0.181  & -0.813 & \multirow{2}{*}{0.691} & -0.180 & -0.795 & \multirow{2}{*}{0.693}  \\
&& (0.0001) & (0.0051) & & (0.0001) & (0.0049) & \\ \cline{2-8}
& \multirow{2}{*}{$\bm{P}_2$} & 0.404 & 1.814 & \multirow{2}{*}{0.309} & 0.405 & 1.793 & \multirow{2}{*}{0.307} \\ 
&& (0.0002) & (0.0016) & & (0.0002) & (0.0016) & \\
\hline
\end{tabular}%
\caption{Estimated random-effects of the JMDF specified in Eq. \eqref{eq:ourJM_appHF} with
Gaussian and Uniform initialization procedures, when $K = 3$ and $K = 2$ mass points are identified. Each point $k$ is reported in terms of its coordinates $\widehat{\bm{P}}_k=\left(\widehat{P}^u_k, \widehat{P}^v_k\right)$ and its weight $\widehat{w}_k$. Standard errors of the estimates are reported in brackets.}
\label{tab:re_sd}
\end{table}

\end{document}